\documentclass[aip,reprint,amsmath,amssymb]{revtex4-1}

\usepackage{graphicx}
\usepackage{dcolumn}
\usepackage{float}
\usepackage{bm}
\usepackage{amssymb}
\usepackage{lipsum}
\usepackage{color}

\begin{document}

\title{Elastic properties of polycatenane chains and ribbons}

\author{James M. Polson} 
\affiliation{ Department of Physics, University of Prince Edward Island, 550 University Avenue, Charlottetown, Prince Edward Island, C1A 4P3, Canada }
\author{Liam MacNevin}
\affiliation{ Department of Physics, University of Prince Edward Island, 550 University Avenue, Charlottetown, Prince Edward Island, C1A 4P3, Canada }
\author{Alaaddin Elobeid}
\affiliation{ Department of Physics, University of Prince Edward Island, 550 University Avenue, Charlottetown, Prince Edward Island, C1A 4P3, Canada }
\author{Carlos E. Padilla Robles}
\affiliation{ Department of Physics, University of Prince Edward Island,
550 University Avenue, Charlottetown, Prince Edward Island, C1A 4P3, Canada }

\date{\today}

\begin{abstract}
Single-chain elasticity is of fundamental importance in polymer physics, as it underlies many of the unique properties of polymer systems. Recently, there has been interest in characterizing the elastic properties of catenanes, molecular architectures composed of linked molecular rings. To date most studies have focused on the force-extension behavior of polycatenane and catenane dimers. In this study, we employ Monte Carlo computer simulations to investigate the elastic properties of a collection of catenane chains. In addition to polycatenane, we also examine the properties of catenane ribbons constructed by connecting two or three polycatenane chains together with a variable number of side-link rings. After first characterizing the behavior of free polycatenane chains and catenane ribbons, we examine their mechanical response to both an elongational force and a torque applied to the end rings of the chain. We find that the stretching induced by the force is counterbalanced by increasing the torque, which tends to twist the chains and in so doing reduce the extension length. At low torque, the twist angle of the end rings of the chain varies linearly with torque, and the associated torsional spring constant, characterizing the resistance of the chain to twist with the applied torque, tends to increase with stretching force. Relative to polycatenane, ribbons tend to be more elongated at low force and less elongated at strong force. In addition, increasing the ribbon width dramatically increases the torsional stiffness of the chain. Finally, decreasing the degree of side-linking in ribbons tends to decrease slightly the extension length at moderate force and to increase the torsional stiffness for sufficiently large gaps.
\end{abstract}

\maketitle

\section{Introduction}
\label{sec:intro}

Mechanically interlocked molecules (MIMs) are permanently linked molecular architectures where two or more components are held together by mechanical bonds. A mechanical bond is defined as ``an entanglement in space between two or more molecular entities... that cannot be separated without breaking or distorting the chemical bonds between atoms''.\cite{Bruns_book}  MIMs have played an important role in the field of molecular switches and molecular machines\cite{Bruns_book, sauvage2017chemical, feringa2017art, stoddart2017mechanically}  and have also been employed in applications ranging from drug delivery\cite{garcia2014cyclodextrin, zhang2013cyclodextrin} to catalysis.\cite{leigh2014rotaxane} Mechanically interlocking polymers (MIPs) are one important class of MIMs that have been the subject of numerous studies, in large part due to their relevance in the design of functional/smart materials\cite{niu2009polycatenanes, hart2021material} as well as their relevance to biological systems.\cite{pieters2016natural, wang2017protein, zhao2018stability} Two notable classes of MIPs are polyrotaxane\cite{huang2005polypseudorotaxanes, harada2009polymeric, arunachalam2014recent, hart2021material} and polycatenane.\cite{niu2009polycatenanes, hart2021material, orlandini2021topological, liu2022polycatenanes}  Polyrotaxane consists of molecular rings threaded through a polymer, along  which they can slide, with bulky stoppers present at the polymer ends to prevent dethreading. By contrast, polycatenanes are chains composed of interlocking molecular rings, with individual rings functioning as monomeric units. The present study is largely concerned with this latter type of MIP.

Recent advances in chemical synthesis methods\cite{wu2017poly, datta2020self} have provided a means to design and construct polycatenanes, whose unique physical properties imparted by the topological bonds can then be characterized using a variety of experimental techniques. This work has in turn prompted numerous simulation studies of polycatenane,\cite{rauscher2018topological, rauscher2020thermodynamics, rauscher2020dynamics, dehaghani2020effects, lei2021dimensional, li2021double, chiarantoni2022effect, tubiana2022circular, caraglio2017mechanical, wu2017poly, chen2023topological, chen2024nonlinear, chiarantoni2023linear, guo2023theta, amici2019topologically} which have sought to illuminate the equilibrium conformational statistics\cite{lei2021dimensional, rauscher2018topological, li2021double,  chiarantoni2022effect, tubiana2022circular, chiarantoni2023linear, dehaghani2020effects, guo2023theta, chen2023topological, chen2024nonlinear} and the dynamics\cite{rauscher2018topological, rauscher2020dynamics, chiarantoni2022effect} of these topologically bound structures under a variety of conditions. For example, the effects of solvent quality,\cite{dehaghani2020effects, guo2023theta} confinement,\cite{amici2019topologically, chiarantoni2023linear} ring-polymer bending rigidity,\cite{chiarantoni2022effect} application of an external stretching force, \cite{wu2017poly, caraglio2017mechanical, chen2023topological, chen2024nonlinear} and supramolecular twist\cite{tubiana2022circular} have all been recently examined.

One interesting aspect of polycatenanes worth exploring is the effect that the mechanical bonds have on their elastic properties, an understanding of which can be facilitated by comparing their observed behavior with that of standard polymers. Single-chain elasticity is of fundamental importance polymer physics as it underlies many of the unique properties of polymeric systems. In recent decades, single-molecule force spectroscopy methods have provided an effective means to characterize the elastic properties of single biopolymers.\cite{bustamante1994entropic, bustamante2003ten, killian2018optical,  bustamante2021optical}  They have been used to generate force-extension curves of tethered molecules to analyze their elastic properties.\cite{bustamante1994entropic, marko1995stretching} They have also been used to investigate the mechanical properties of nucleic acids,\cite{smith1996overstretching, wang1997stretching} peptides,\cite{tskhovrebova1997elasticity, kellermayer1997folding} chromatin fibres,\cite{cui2000pulling, bennink2001unfolding} and nucleoprotein filaments.\cite{hegner1999polymerization}  An extension of  standard optical tweezers, the angular optical trap (AOT) can be used to investigate the torsional stiffness of biomolecules. An AOT can simultaneously measure the force, position, torque and angle of a trapped birefringent particles such as a nanofabricated quartz cylinder.\cite{forth2013torque, la2004optical, deufel2007nanofabricated, inman2010passive}  It has been used to characterize physical states and map out the force-torque phase diagram for DNA.\cite{forth2008abrupt, daniels2009discontinuities, sheinin2009twist,gao2021torsional, forth2013torque, gao2022angular} As a result of this work as well as numerous complementary theoretical and computer simulation studies, an understanding of elasticity of single polymers is now well advanced.\cite{saleh2015perspective, camunas2016elastic}

To the best of our knowledge, no such single-molecule experiments to study the elasticity of polycatenane or any other MIP have yet been carried out. However, in anticipation that such experiments will eventually be done, some researchers have carried out computer simulation studies that focus on the stretching behaviour  of polycatenane chains subject to an applied force.\cite{wu2017poly, caraglio2017mechanical, chen2023topological, chen2024nonlinear} For example, Wu {\it et al.} used  molecular dynamics (MD) simulations to study an all-atom model of a metallated polycatenane in an explicitly modelled solvent.\cite{wu2017poly} A stretching force was applied by connecting the polycatenane through the periodic boundaries of the box and systematically controlling the box length. They found that the resulting force-extension curve was well described using the extensible wormlike chain model.\cite{odijk1995stiff} Caraglio {\it et al.}\cite{caraglio2017mechanical} carried out Langevin dynamics simulations to study a two-linked-ring system with semi-flexible rings modelled using the Kremer-Grest (KG) model.\cite{kremer1990dynamics} They found that the average extension of linked rings, once normalized with respect to a single ring of equivalent contour length, is not monotonic in the applied force, a feature they attribute to different stretching compliance of the linked portion and the rest of the rings contour. More recently, Chen {\it et al.}\cite{chen2023topological} carried out Langevin dynamics simulations to study polycatenane with fully-flexible KG-model rings. They found that in the linear-elasticity (i.e., low-force) regime, topological catenation leads to an enhancement in the elastic modulus of polycatenane compared with that for a bonded-ring counterpart system in which the rings are connected with regular (i.e., covalent-like) bonds as opposed to mechanical bonds.\cite{chen2023topological} In a follow-up study,\cite{chen2024nonlinear} Chen {\it et al.} further examined the same system, this time focussing on the non-linear elasticity regime. Here, they observe three sub-regimes, two of which exhibit stress-stiffening behavior, and between which lies a stress-softening regime. The latter regime is was not observed in bonded-ring counterpart system. They attributed this regime to ``rotational sliding'', wherein of one ring slides around its catenated neighbor as is simultaneously changes orientation, a process that can only occur if the catenated ring is sufficiently oblate and has a sufficiently great contour length. 

In addition to polycatenane, other catenated structures have attracted considerable attention in recent years. One of the most interesting examples is the kinetoplast, a complex DNA structure found in the mitochondria of trypanosome parasites. It is often described as ``molecular chainmail", as it consists of thousands of circular DNA molecules, known as minicircles, as well as a smaller number of larger rings called maxicicles, topologically linked in a 2D network.\cite{michieletto2025kinetoplast} As polycatenane is an analogue to 1D polymers, so too is a kinetoplast an analogue to 2D polymers. The unique physical properties of kinetoplasts have been the subject of much research in the past few years.\cite{klotz2020equilibrium, soh2020deformation,  soh2020ionic, soh2021equilibrium, yadav2021phase,  yadav2023tuning, he2023single, diggines2024multiscale, ramakrishnan2024single} Their elastic properties have been examined indirectly using measurements with nanofluidic techniques\cite{klotz2020equilibrium} and, more recently, atomic force microscopy.\cite{ramakrishnan2024single, diggines2024multiscale} These studies suggest that their bending stiffness is two orders of magnitude smaller than lipid bilayer membranes. In addition, changing the network topology by cleaving minicircles with restriction enzymes was observed to change the geometric properties and morphology of the network and significantly reduce the bending rigidity.\cite{ramakrishnan2024single}
Between the 1D-polymer analogue of polycatenane and the 2D-polymer analogue of kinetoplasts lies the as yet unexplored domain of catenated {\it ribbons}. While the conformational properties of ribbon-like molecules have been the subject of theoretical and computational studies over the past three decades,\cite{nyrkova1996highly, everaers1995fluctuations, golestanian2000statistical, mergell2002statistical, arinstein2005conformational, michaels2023conformational} the properties of ribbon catenanes have yet to be investigated.

In this study, we use Monte Carlo (MC) computer simulations to investigate the elastic properties of polycatenane chains and ribbons. In addition to measuring the variation of the extension with respect to a stretching force as in previous studies of polycatenane, we also characterize the torsional elastic properties. By analogy with experiments carried out on biomolecules such as DNA with AOTs, this is done by applying a torque to the end rings of the chain,\cite{forth2008abrupt, daniels2009discontinuities, sheinin2009twist, gao2021torsional, forth2013torque, gao2022angular} which are themselves tethered to hard surfaces. For computational efficiency, we use catenanes composed of rigid rings, leaving the case of flexible and semi-flexible chains for future work. As most previous studies of polycatenane have employed models with flexible or semi-flexible rings, we first characterize the standard scaling properties of the rigid-ring chains. In the case of ribbons, we also investigate the effects of creating gaps by removing some rings in the network, by analogy with experiments on kinetoplasts where ring polymers are removed using restriction enzymes.\cite{ramakrishnan2024single} We find that force-extension behavior in the absence of a torque is comparable to that observed in previous studies using fully-flexible and semi-flexible chains. Application of a torque twists the chains in a manner that tends to oppose the extensional force and reduce the extension of polycatenane and catenane ribbons. At low torque, the twist angle between the end rings varies linearly with torque, and the corresponding torsional spring constant increases monotonically with both the applied force and with the ribbon width. Finally, the introduction of gaps in ribbons tends to decrease the extension length at moderate forces and to increase the torsional stiffness when the gaps are sufficiently large.

The remainder of the article is organized as follows. Section~\ref{sec:modelmethod} describes the model employed in the simulations and also describes in detail the MC procedure used to carry out the simulations. Section~\ref{sec:results} presents the results of the calculations. Finally, Sec.~\ref{sec:conclusions} summarizes the main conclusions of this work.

\section{Model and Methods}
\label{sec:modelmethod}

\begin{figure}[!ht]
\begin{center}
\vspace*{0.1in}
\includegraphics[width=0.45\textwidth]{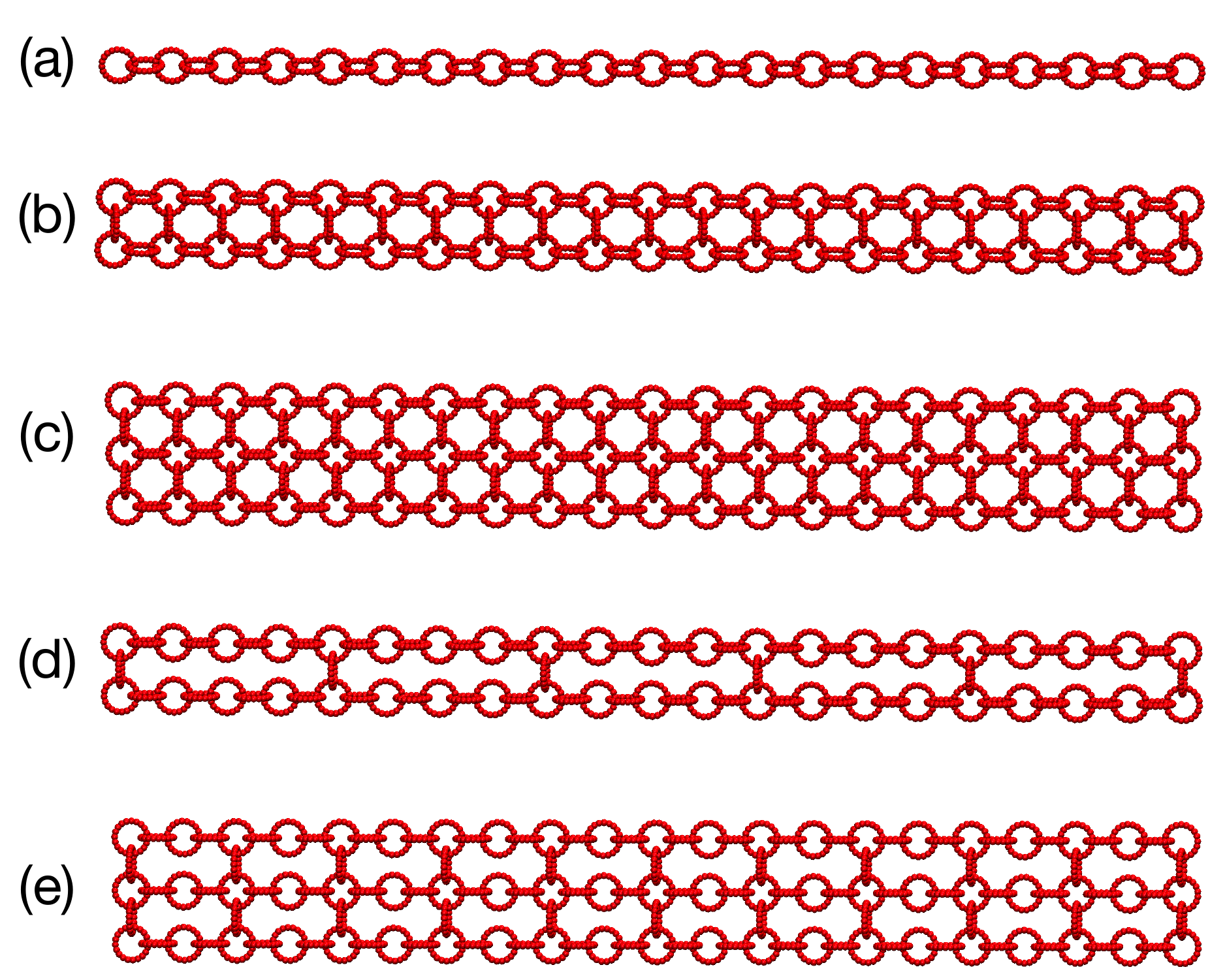}
\vspace*{0in}
\end{center}
\caption{Illustration of the catenated chains examined in this study. Each chain in the illustration has a length of $N$=41. (a) Polycatenane ($n_{\rm w}=1$). (b) Catenane ribbon of width $n_{\rm w}$=2. (c) Catenane ribbon of width $n_{\rm w}$=3. (d) Catenane ribbon of width $n_{\rm w}$=2 and a gap length of $g$=4. (e) Catenane ribbon of width $n_{\rm w}$=3 and a gap length of $g$=2. }
\label{fig:illustration1}
\end{figure}

MC simulations are used to study chains composed of interlocking rigid  circular rings of diameter $D$. Each ring is composed of a set of $m=20$ tangentially connected hard spheres of diameter $\sigma$, which is essentially the thickness of the ring.  
We examine catenated structures of various topologies, each illustrated in Fig.~\ref{fig:illustration1}. The simplest structure is a polycatenane chain, shown in Fig.~\ref{fig:illustration1}(a). In this case, the chain linkage is sequential, with  each ring linked with two other rings, with the exception of the end rings. Catenane ribbons are constructed by linking two or more polycatenane chains every second ring. We define the ribbon width, $n_{\rm w}$, as the number of chains linked in this way, with $n_{\rm w}=1$ corresponding to the polycatenane chain shown in Fig.~\ref{fig:illustration1}(a). Fig.~\ref{fig:illustration1}(b) shows a ribbon of width $n_{\rm w}=2$, and Fig.~\ref{fig:illustration1}(c) shows a ribbon constructed using $n_{\rm w}=3$ chains. Note the square linking architecture for the chains in Figs.~\ref{fig:illustration1}(b) and (c). In the case of ribbons, we also consider architectures in which some of the side-link rings connecting the polycatenane chains are removed. We define the gap parameter $g$ such that only every $g$th side link remains after removing others. As an illustration, Fig.~\ref{fig:illustration1}(d) shows a ribbon with $n_{\rm w}$=2 and a gap parameter of $g$=4. Likewise, Fig.~\ref{fig:illustration1}(e) shows a ribbon with $n_{\rm w}$=3 and a gap parameter of $g$=2. Note that $g$=1 corresponds to no gaps in the side links, as in the ribbons in Figs.~\ref{fig:illustration1}(b) and (c). In the remainder of the article the label ``chain'' refers to any of the structures illustrated in the figure, ``polycatenane'' refers specifically to the case of $n_{\rm w}=1$ in Fig.~\ref{fig:illustration1}(a) (i.e., its usual definition), and ``ribbons'' refer specifically to the case of $n_{\rm w}\geq 2$. 

\begin{figure}[!ht]
\begin{center}
\vspace*{0.1in}
\includegraphics[width=0.42\textwidth]{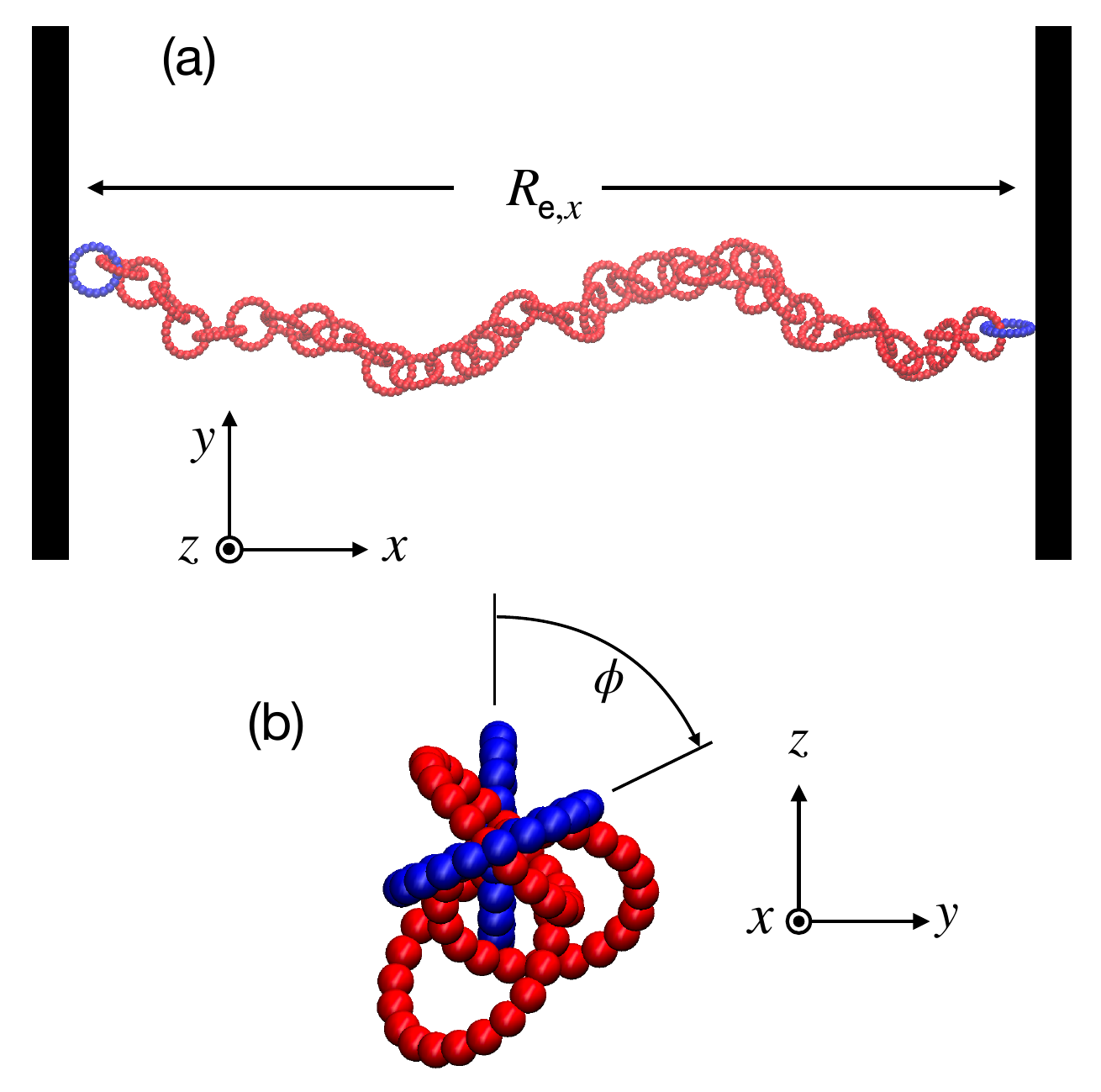}
\vspace*{0in}
\end{center}
\caption{(a) Snapshot of a polycatenane chain of length $N=39$ with end rings tethered to two parallel hard walls. The end rings are bound to the walls in a perpendicular orientation. The extension length, $R_{{\rm e},x}$, i.e. the distance between the centers of the end rings along the $x$ axis, is labelled. (b) Illustration of the definition of the twist angle, $\phi$, of the end rings. For visual clarity, the end rings shown in the snapshot have the same $y$ and $z$ coordinates, though this is not a constraint imposed in the simulations.}
\label{fig:illustration2}
\end{figure}

In most simulations, the rings at each end of the catenane chains and ribbons are tethered two hard, flat, parallel walls, between which the chains are confined. At each end of the chain or ribbon, $n_{\rm w}$ end rings are fixed perpendicular and tangential to the hard wall. This is illustrated in Fig.~\ref{fig:illustration2}, which shows a snapshot of a polycatenane chain ($n_{\rm w}=1$) bound to two walls. In this case, just a single end ring is shown attached to each wall. In the case where $n_{\rm w}\geq 2$, the $n_{\rm w}$ end rings are held at fixed positions and orientations along the wall, such that the planes of the end rings align with the displacements between them. The position and orientation of one wall is held fixed, while the other wall can move along the $x$ axis, as depicted in the figure. This means that the distance between the walls, and thus the distance between the end rings of the chain can vary. We call the distance between the centers of the end rings at opposite ends of the chain along the $x$ axis the extension length, $R_{{\rm e},x}$. The moveable wall can also rotate around the $x$ axis in a manner that causes the attached end rings to rotate. The angle between end rings at opposite ends of the chain is called the twist angle, $\phi$. Figure~\ref{fig:illustration2}(b) illustrates the twist angle for the case of polycatenane ($n_{\rm w}=1$). Note that since the end-ring orientation can only change by rotations about the $x$ axis, the two walls remain parallel. Finally, note the moveable wall and the end rings tethered to it can also translate in the $y-z$ plane.  

The tethered chains and ribbons are stretched and twisted by applying a constant external force, $f$, and torque, $\tau$, to the end rings. To apply the force, the end ring(s) on the moveable wall are subject to a potential energy given by
\begin{eqnarray}
U_{\rm stretch} = - f x,
\label{eq:Ustretch}
\end{eqnarray}
where $x$ is the coordinate of the end ring(s) on the moveable wall. Likewise, the torque is applied by imposing the potential energy given by
\begin{eqnarray}
U_{\rm twist} = - \tau\phi,
\label{eq:Utwist}
\end{eqnarray}
where $\phi$ is the twist angle defined above and illustrated in Fig.~\ref{fig:illustration2}(b).

In addition to simulations employing the model described above, we also carry out some simulations for systems without confining walls.  For these wall-less systems, we consider free chains as well as stretched chains. However, in the latter case, no torque applied to the end rings. 

Note that tethering the chains to walls serves two purposes. First, it mimics the systems studied in single-molecule force spectroscopy experiments, in which a biomolecule such as DNA is tethered to a hard surface at one end and to an optical bead at the other end. The second purpose is more practical. In order to twist the chain by applying a torque, tethering to surfaces at each end is required in order to sustain the twist and prevent unwinding by the sliding of rings linked to the end rings around those end rings. Note that other means can be used to twist the chains. For example, in Ref.~\onlinecite{tubiana2022circular} supramolecular twist was maintained without such wall tethering by connecting the end rings of polycatenane together to form a ``circular polycatenane'', i.e., a catenane analogue of a ring polymer. However, such structures are not relevant to the present work, which seeks to characterize the response of the chain twisting to an applied torque. 

The MC simulations use the standard Metropolis methodology. For convenience, the initial positions and orientations of the rings are chosen to  correspond the configurations shown in Fig.~\ref{fig:illustration1}. Two types of MC trial moves are carried out for randomly selected rings: (1) random displacement, and (2) random rotation of the normal axis of the ring about a randomly chosen axis. In the first move, all spherical beads are moved by equal amounts, and in the second move, the positions of the beads on the rotated circular ring are also randomly replaced along the ring. 

Trial moves are tested to see whether the new configuration preserves the original linking structure, i.e., rings that are originally linked must stay linked, and rings that were originally unlinked must remain so. Any move that violates these constraints is rejected. In addition, the move is rejected if any pair of spheres on different rings overlap. Moves that preserve the link structure and do not result in such overlap are accepted with a probability of ${\rm min}(1,\exp(-\Delta U/k_{\rm B}T))$, where $\Delta U$ is the difference in energy between old and new configurations calculated from the stretching and twisting potentials of Eqs.~(\ref{eq:Ustretch}) and (\ref{eq:Utwist}). Note that these potentials only involve the coordinates of the end rings, and thus $\Delta U=0$ for the moves of all rings other than the end rings.  Maximum displacement and rotation angles are chosen to yield an acceptance ratio in the range 30 -- 50\%. Displacement and rotation moves are selected with equal probability. A MC cycle is defined as a sequence of $N_{\rm rings}$ consecutive trial moves, where $N_{\rm rings}$ is the total number of rings in the chain. Thus, during each cycle {an attempt is made to either translate or rotate each ring once,} on average. Prior to data sampling, the system is equilibrated for a period chosen to ensure the complete decay of the initial transients in the histories of all measured quantities. Equilibration periods were typically between $10^6$ and $10^7$ MC cycles, and production runs were in the range of $5\times 10^6$ to $1\times 10^8$ MC cycles in duration. Typically, the longer equilibration and production periods were required for longer polycatenane chains and for the catenane ribbons with $n_{\rm w}$=2 and  $n_{\rm w}$=3. The results of between 80 and 200 independent simulations were averaged to achieve reasonable statistical accuracy, with larger systems requiring more averaging.

In the results presented below, distances are measured in units of $D$, energy
is measured in units of $k_{\rm B}T$, force is measured in units of $k_{\rm B}T/D$
and torque is measured in units of $k_{\rm B}T$.

\section{Results}
\label{sec:results}

\subsection{Free chains}
\label{sec:free_chains}

\begin{figure}[!ht]
\begin{center}
\vspace*{0.0in}
\includegraphics[width=0.45\textwidth]{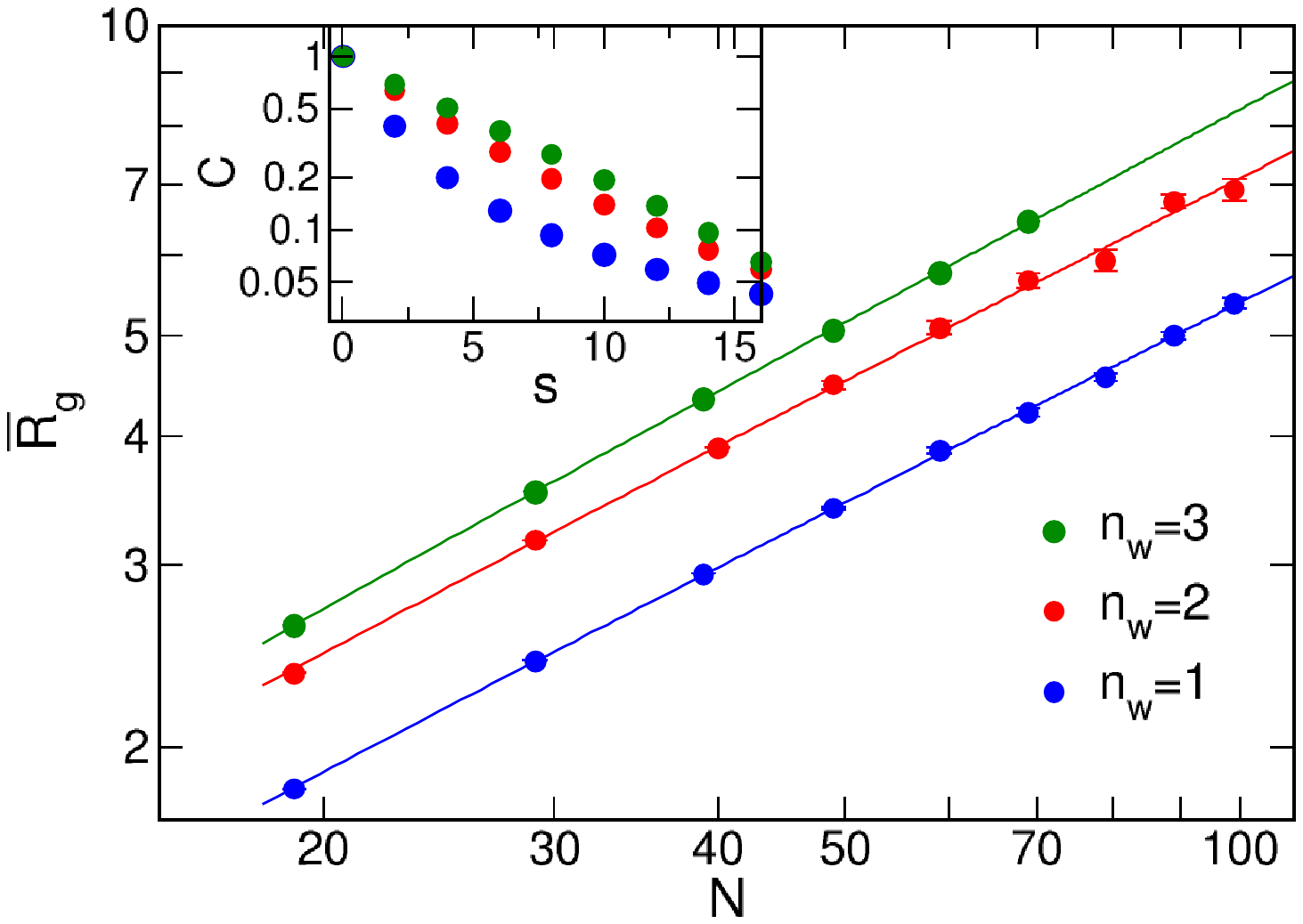}
\vspace*{-0.2in}
\end{center}
\caption{RMS radius of gyration of catenane chains vs chain length, $N$. Results are shown for chains of width $n_{\rm w}=1$, $n_{\rm w}=2$, and $n_{\rm w}=3$. The solid lines show fits to the power function $\bar{R}_{\rm g}=cN^\nu$. The fits yielded scaling exponents of $\nu=0.65\pm 0.01$ for $n_{\rm w}=1$, $\nu=0.66\pm 0.01$ for $n_{\rm w}=2$, and $\nu=0.696\pm 0.003$ for $n_{\rm w}=3$. The inset shows the tangent-tangent correlation function $C(s)$ for chains of length $N=39$ and width $n_{\rm w}=1$, $n_{\rm w}=2$ and $n_{\rm w}=3$.}
\label{fig:rg_vs_n1}
\end{figure}

We consider first the case of free catenane chains and characterize the scaling of the physical size of the chains with respect to the chain length and width. As a measure of size, we use the root-mean-square (RMS) radius of gyration,  $\bar{R}_{\rm g}\equiv \sqrt{\langle R_{\rm g}^2 \rangle}$, where the radius of gyration of a given configuration is defined
\begin{equation}
R_{\rm g} \equiv \frac{1}{N_{\rm rings}} \sum_{n=1}^{N_{\rm rings}} \left | \vec{r}_n - \vec{r}_{\rm cm} \right |^2,
\label{eq:rgdef}
\end{equation}
where 
$\vec{r}_n$ is the position of the center of the $n$th ring, and $\vec{r}_{\rm cm}$ is the effective center-of-mass position associated with the coordinates of the ring centres, i.e. $\vec{r}_{\rm cm}\equiv (1/N_{\rm rings})\sum_{n=1}^{N_{\rm rings}} \vec{r}_{n}$. Figure~\ref{fig:rg_vs_n1} shows the variation of $\bar{R}_{\rm g}$ with respect to chain length, $N$. Results are shown for chains of width $n_{\rm w}=1$, $2$ and $3$. The solid lines are fits of the data to the power function $cN^\nu$, which yield scaling exponents of $\nu=0.65\pm 0.01$ for $n_{\rm w}=1$, $\nu=0.66\pm 0.1$ for $n_{\rm w}=2$, and $\nu=0.696\pm 0.003$ for $n_{\rm w}=3$. These exponents are somewhat larger than the Flory exponent for a self-avoiding chain of $\nu_{\rm F}\approx 0.588$.\cite{Rubenstein_book} 

It is likely that finite-size effects contribute to the discrepancy between the measured scaling exponents and the Flory value. In the case of polycatenane at least, it is expected that the exponent would converge to $\nu_{\rm F}$ in the limit of large $N$. Other simulation studies have reported values comparable to ours for the scaling exponent of polycatenane. Note that these simulations have typically use the Kremer-Grest (KG) model\cite{kremer1990dynamics} to describe the rings,  which differs from but is comparable to the hard-sphere model employed in the present work. For example, Dehaghani {\it et al.} found $\nu\approx 0.63$ for chains composed of semi-flexible rings (with ring persistence length to contour length ratio of 5) for comparable ring contour lengths used here ($m=20$).\cite{dehaghani2020effects} Chiarantoni and Micheletti report a value of $\nu\approx 0.64$ for chains with rings with various bending rigidities and contour lengths.\cite{chiarantoni2022effect} Li {\it et al.} quantified the chain-length dependence of the effective exponent for chains composed of fully-flexible rings (see Eq.~(4) of  Ref.~\onlinecite{li2021double}), the results of which yield values ranging from $\nu\approx 0.65$ for $N=19$ to $\nu\approx 0.614$ for $N=99$ (i.e., the range of $N$ used to fit the data in in Fig.~\ref{fig:rg_vs_n1}). The values of $\nu$ for $n_{\rm w}=1$ and 2 extracted from the fits in Fig.~\ref{fig:rg_vs_n1} are close to but slightly higher than those previously reported values for flexible-ring and semiflexible-ring models. This difference may be due to using rigid rings in the present model. The value for $n_{\rm w}=3$ is significantly higher than those for the other two cases. In this case, the ribbon-like conformational behavior likely emerges,\cite{michaels2023conformational}  with a greater stiffness of the ``turning'' and ``twisting'' modes of the ribbons relative to the undulation mode (see Fig.~1 of Ref.~\onlinecite{michaels2023conformational}). (A significant increase in the twisting mode stiffness with increasing ribbon width is demonstrated in Sec.~\ref{subsec:ribbons}.) In the limit of increasing stiffness of these modes, an ideal ribbon behaves like a 2D wormlike chain.\cite{michaels2023conformational} As a self-avoiding 2D chain has a scaling exponent of $\nu=0.75$, it is unsurprising that the scaling exponent tends toward this value as the ribbons widen.

The other key trend in Fig.~\ref{fig:rg_vs_n1} is the increase of $\bar{R}_{\rm g}$ with increasing ribbon width, $n_{\rm w}$. For example, increasing the ribbon width from $n_{\rm w}=1$ to $n_{\rm w}=2$ increases $R_{\rm g}$ by a factor very close to 1.3 for all values of $N$. Likewise, increasing the width from $n_{\rm w}=1$ to $n_{\rm w}=3$ increases $R_{\rm g}$ by a factor that increases from 1.44 for $N=19$ to 1.53 for $N=69$. This increase is likely due to a combination of an increase in excluded-volume interactions as the ribbon thickens as well as an entropic stiffening of the chain as the number of topological linkages increase with $n_{\rm w}$. To quantify the latter effect, we calculate the tangent-tangent correlation of the mechanical bonds at different separation $s$ along the chain:
\begin{eqnarray}
C(s) = \left\langle \hat{r}_i \cdot \hat{r}_{i+s}\right\rangle,
\label{eq:Cs}
\end{eqnarray}
where $\hat{r}_i$ is the unit vector pointing from the center of ring $i$ to that of ring $i+1$. Note that $i$ is an integer index for the rings along the contour of the polycatenane chain or else along each side-linked polycatenane chain in a ribbon. In addition, the average is taken over all sampled conformations  and by sliding the ring index $i$ over the permitted range. In the case where $n_{\rm w}\geq 2$, bonds formed from the side-link rings are excluded from the calculation, and the quantity is further averaged over the $n_{\rm w}$ linked polycatenane chains that comprise the ribbon. The inset of Fig.~\ref{fig:rg_vs_n1} compares $C(s)$ calculated for $n_{\rm w}=1$, $2$ and $3$.  The effective persistence length $p$  is defined by the relation by the relation $C(p)=1/e$.  Note that $p$ is dimensionless as it quantifies the number of mechanical links over which the directionality of the bond vectors decays. This analysis yields $p_1\approx 2.2 $, $p_2\approx 4.6$ and $p_3=5.9$, where $p_i$ denotes for the values $p$ for $n_{\rm w}=i$. This increase in $p$ can be used with some basic results of polymer theory to account for the corresponding increase in $R_{\rm g}$. Recall that a typical semiflexible polymer has a size that scales as $R_{\rm g}\sim l_{\rm K}(w/l_{\rm K})^{2\nu-1}(L_{\rm c}/l_{\rm K})^\nu$, where $L_{\rm c}$ is the contour length of the chain, $l_{\rm K}$ is the Kuhn length, and $w$ is the thickness of the polymer.\cite{Rubenstein_book} To apply this standard result to the catenated chains, we assume that the ratio $l_{\rm K}/p$ is unaffected by $n_{\rm w}$ (i.e., negligible change in the RMS distance between linked rings along the chain). It follows that $R_{\rm g}^{(2)}/R_{\rm g}^{(1)} = (l_{\rm K,2}/l_{\rm K,1})^{2-3\nu}(w_2/w_1)^{2\nu-1}$, where the subscripts and superscripts again denote the value of $n_{\rm w}$. Using $l_{\rm K,2}/l_{\rm K,1}=p_2/p_1=2.1$, choosing $w_2/w_1=2$, and using the measured scaling exponent for $n_{\rm w}=1$ of $\nu=0.65$ yields $R_{\rm g}^{(2)}/R_{\rm g}^{(1)} =1.3$, in accord with the measured value. The perfect agreement of the prediction with the measurement may be a result of a fortuitous cancelation of errors in the approximation, which includes the reasonable but somewhat arbitrary choice of $w_2/w_1=2$. Note that the exponent value of $2-3\nu=0.05$ suggests that the increase in the effective persistence length has negligible impact on $R_{\rm g}$, whose increase appears to be mainly due to increase in the excluded-volume interactions associated with thickening the ribbon.

Note that the value of $p$ for polycatenane is somewhat larger than the value of $p\approx 1.8$ reported in Ref.~\onlinecite{chiarantoni2022effect} for a comparable molecular model with chains composed of semi-flexible rings of length $m=20$ beads and at high ring bending rigidity. However, using the alternative definition of $p$ employed in  Ref.~\onlinecite{chiarantoni2022effect}, i.e., $p\equiv\int_0^\infty C(s)\,ds$, and carrying out a numerical evaluation yields a value of $p\approx 1.83$, in much better agreement with those results. Finally, note that for the case of polycatenane at least, the effective persistence length arises from excluded volume (both topological excluded volume\cite{des1981ring} and the standard variety) between sequentially separated rings rather than from just interactions between linked rings.\cite{chiarantoni2022effect}

\subsection{Unconfined polycatenane under tension}
\label{subsec:polycat_stretch}

\begin{figure}[!ht]
\begin{center}
\vspace*{0.0in}
\includegraphics[width=0.45\textwidth]{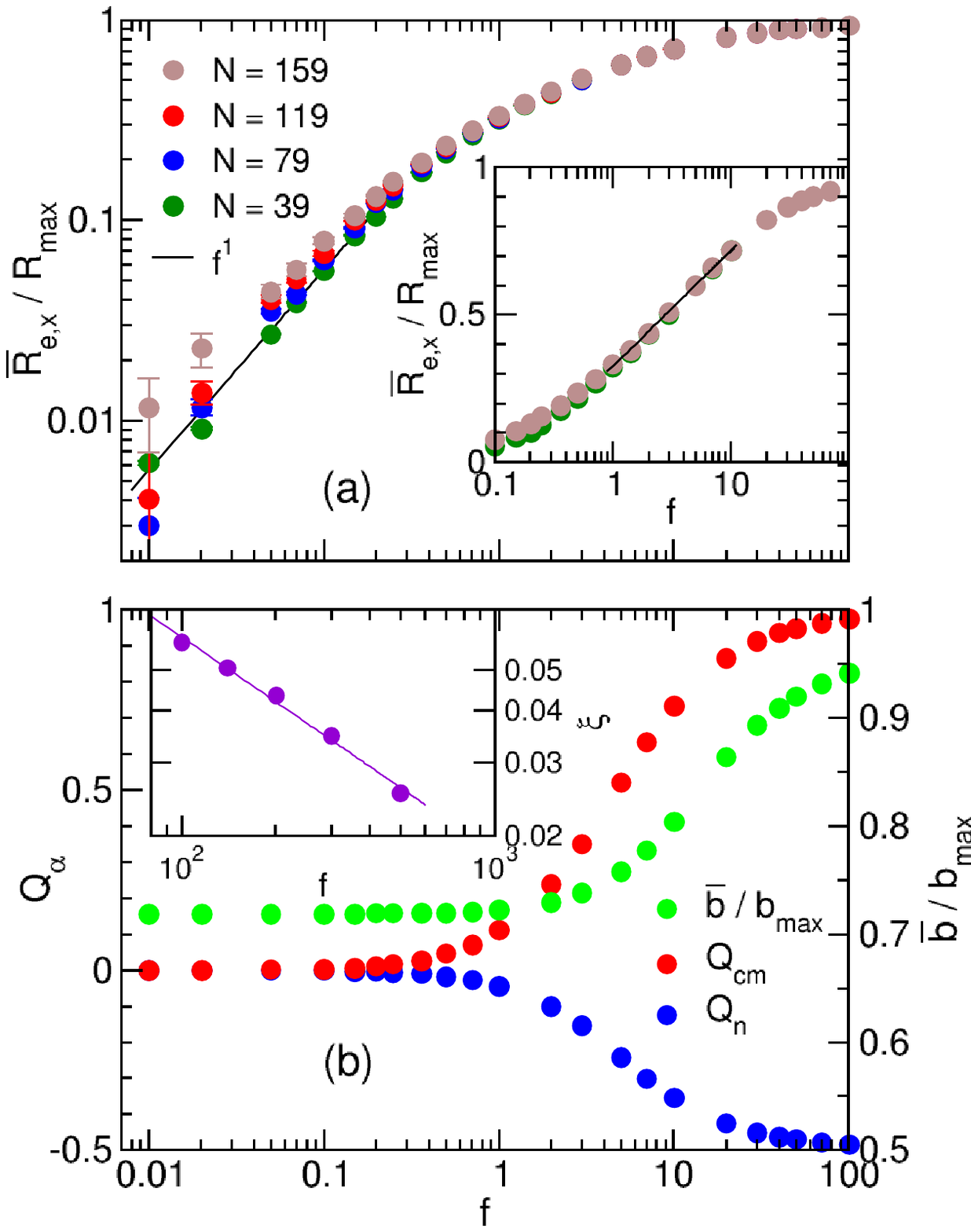}
\vspace*{-0.2in}
\end{center}
\caption{(a) Scaled mean extension length, $\bar{R}_{{\rm e},x}/R_{\rm max}$, vs force applied to the end rings for a polycatenane chain. No confining walls are present, and no torque is applied to the end rings. Results for various values of the chain length, $N$, are shown. The solid line shows the relation $f^{1}$, the scaling expected in the weak force limit for standard polymers absent any confinement. The inset shows the data for $N=159$ and $N=39$ plotted on a semi-log scale. The solid line is a fit of the $N=159$ data to the function $c_0+c_1\ln(f)$ in the range $1\leq f \leq 10$. 
(b) Orientational order parameters $Q_{\rm cm}$ and $Q_{\rm n}$ (defined in the text) and scaled RMS distance between linked rings $\bar{b}/b_{\rm max}$ vs $f$ for the case of $N=159$. The inset shows $\xi\equiv 1-\bar{b}/b_{\rm max}$ vs $f$ in the high-force limit of $f\geq 100$. The solid line is a fit to the data using a power function $\xi=cf^{-\alpha}$ with $\alpha=0.52\pm 0.03$ and $c=0.66\pm 0.09$. }
\label{fig:rex_Q_freeends}
\end{figure}

Next, we narrow the focus to the case of polycatenane chains, illustrated in Fig.~\ref{fig:illustration1}(a). We first consider the effects of stretching the chain with an applied force but with no applied torque and in the absence of confining walls. Figure~\ref{fig:rex_Q_freeends}(a) shows the variation of the relative mean extension length, $\bar{R}_{{\rm e},x}/R_{\rm max}$, where $R_{\rm max}$ is defined  as the maximum extension length of each chain permitted by the linking and non-overlap constraints of the model. Results are shown for the four different chain lengths of $N$=39, 79, 119 and 159. As expected, the extension length increases monotonically with respect to $f$. The results all converge for $f\gtrsim 1$, while in the case of $f\lesssim 1$, the trend appears to be that $R_{{\rm e},x}$ is increases slightly with increasing $N$. 

Overlaid on the simulation data for low force is the power function $f^1$, the scaling expected for chain extension  in the case of standard polymers in the limit where $fR_{\rm F}/k_{\rm B}T\ll 1$, where $R_{\rm F}$ is the Flory radius of a free polymer chain.\cite{deGennes_book} The simulation results for $\bar{R}_{{\rm e},x}$ exhibit a scaling close to this prediction, with a small deviation in the expected scaling coefficient, which decreases slightly from unity as the chain length increases. For example, a fit to the data for $N=159$ in the range $f\leq 0.1$ to the function $f^\alpha$ yields an exponent of $\alpha=0.81\pm 0.03$, whereas the same fit yields $\alpha=1.00\pm 0.07$ for $N=39$. Similar scaling was observed in Ref.~\onlinecite{chen2024nonlinear} for stretching polycatenane comprised of semi-flexible chains. 

The inset of Fig.~\ref{fig:rex_Q_freeends}(a) shows the data for $N=159$ and $N=39$ plotted on a semi-log scale. The solid line shows a fit of the data to $c_0+c_1\ln(f)$ in the range $1\leq f\leq 10$ and demonstrates that $\bar{R}_{{\rm e},x}$ varies linearly with $\ln(f)$ in this range of force. This is consistent with an elastic modulus, $E\equiv df/d\bar{R}_{{\rm e,x}}$, that varies as $E\sim f^\alpha$, where the exponent has a value of $\alpha=1$. As the linear scaling of $\bar{R}_{{\rm e},x}$ with $\ln(f)$ is unaffected by the chain length (the inset shows results for $N=39$ and $N=159$), so too is the linear scaling of $E$ with $f$ invariant with respect to chain length. This result for intermediate force  was also observed in the case of flexible-ring chains in Ref.~\onlinecite{chen2024nonlinear} for   chains comprised of a sufficiently large number of rings. However, in that study, the scaling exponent $\alpha$ tended to decrease with a decreasing number of rings in the range $N\leq 20$. 

Notably absent in the results for rigid-ring chains of this study or those for flexible-ring chains in Ref.~\onlinecite{chen2024nonlinear} is the Pincus regime, where the extension length scales as $\bar{R}_{{\rm e},x} \sim f^{(1-\nu)/\nu}\approx f^{0.70}$ for a Flory exponent of $\nu\approx 0.588$.\cite{Rubenstein_book} In that regime,  the chain stretches out into a sequence of blobs within each of which Flory scaling holds. One regime that was observed in Ref.~\onlinecite{chen2024nonlinear} for polycatenane chains (but not the bonded-ring counterparts) is a stress-softening regime, lying between between two stress-stiffening regimes at intermediate and high forces, respectively. Here, the elastic modulus is constant with respect to the force. Such a regime is not observed in the present results for any chain length. Chen {\it et al.} attributed  this interesting behavior to ``rotational sliding", a process in which a ring tends to slide around a catenated ring while it simultaneously changes orientation as the force increases. They found that such a process is possible if the catenated ring is sufficiently oblate and if the ring length, $m$,  is sufficiently large to permit ring sliding.\cite{chen2024nonlinear} They also note that the low-force boundary of the regime occurs at the onset of an oblate-to-prolate ring shape transition, and though it was noted that this transition was not itself a {\it sufficient} condition for the regime,  it well may be a necessary condition. As our simulations employ rigid rings, there is no oblate-prolate transition, and its absence may well explain the absence of the stress-softening regime.

To further investigate the force-extension behavior, we have calculated quantities that characterize the orientation of the rings. The first quantity is defined
\begin{eqnarray}
Q_{\rm cm} \equiv  \left\langle {\textstyle\frac{3}{2}}\cos^2\theta_{\rm cm}-{\textstyle\frac{1}{2}}\right\rangle,
\end{eqnarray}
where $\theta_{\rm cm}$ is the angle between the displacement vectors connecting the centers of pairs of linked rings and the $x$ axis, along which the force is extended. Likewise, the second quantity is defined
\begin{eqnarray}
Q_{\rm n} \equiv  \left\langle {\textstyle\frac{3}{2}}\cos^2\theta_{\rm n}-{\textstyle\frac{1}{2}}\right\rangle,
\end{eqnarray}
where $\theta_{\rm n}$ is the angle between the normal vector of each ring with respect to the $x$ axis. A third quantity of note is $\bar{b}$, the RMS distance between the centers of pairs of linked rings. 

Figure~\ref{fig:rex_Q_freeends}(b) shows the variation of $Q_{\rm cm}$, $Q_{\rm n}$ and $\bar{b}$ with respect to the stretching force for the case of a chain of length $N=159$. At low force ($f\lesssim 0.5$), both $Q_{\rm cm}$ and $Q_{\rm n}$ are close to zero, indicating a roughly uniform distribution of center-of-mass vectors and ring normal vectors. Likewise, in this range of force, the center-to-center distance between linked rings is roughly constant. This collection of results show that small forces have minimal effect on the conformational behavior of the chain. Above a somewhat higher force of $f\approx 1$, $Q_{\rm cm}$ rises rapidly and by $f=50$ levels off to its maximum value of $Q_{\rm cm}=1$, at which point the inter-ring vectors are aligned with the force direction. Concomitantly, $Q_{\rm n}$ initially  decreases and then eventually levels off to its minimum possible value of $Q_{\rm n}=-0.5$, at which point the ring normals are all perpendicular to the force direction. The figure also shows that the inter-ring distance $\bar{b}$ rapidly increases as the chain stretching increases, eventually levelling off to its maximum value. Note that the range of force over which the mean extension scales linearly with $\ln(f)$ in Fig.~\ref{fig:rex_Q_freeends}(a) (equivalently, where the elastic modulus scales as $E\sim f^1$), i.e. $1\lesssim f \lesssim 10$, is also the regime in which $\bar{b}$ rises most rapidly in Fig.~\ref{fig:rex_Q_freeends}(b). It would be interesting to determine if and how the  observed scaling of $\bar{R}_{{\rm e},x}$ (and therefore $E$) with $f$ arise from extensibility of the chain associated with increase in $\bar{b}$, as well as from the orientational ordering of the rings evident in the rapid change in $Q_{\rm n}$ and $Q_{\rm cm}$ in this regime. Further, does this scaling at intermediate force persist for much longer chains, or is the apparent scaling $E\sim f^\alpha$ with $\alpha=1$ only valid over a restricted range of $N$? (Note that Chen {\it et al.} observed a dependence of $\alpha$ won $N$ for very short chains of length $N\leq 20$.)\cite{chen2024nonlinear} Unfortunately, such an investigation would likely require calculations with chain lengths that are too large to be feasible currently.

The inset of Fig.~\ref{fig:rex_Q_freeends}(b) shows the variation of $\xi\equiv 1-\bar{b}/b_{\rm max}$ vs $f$ in the high-force limit of $f\geq 100$. Overlaid on the data is a fit to the power function $\xi=cf^{-\alpha}$, which yielded values of $\alpha=0.52\pm 0.03$ and $c=0.66\pm 0.09$. This observed scaling accords well with the predictions for the extension length at high force predicted by the Marko-Siggia model for wormlike chains.\cite{marko1995stretching} In the high-force limit of that model, $\bar{R}_{{\rm e},x}/R_{\rm max} \approx 1- \sqrt{k_{\rm B}T/2fl_{\rm K}}$, where $l_{\rm K}$ is the Kuhn length of the polymer. To apply this relation to  polycatenane, we note that $R_{\rm max}=(N-1)b_{\rm max}$ and also that $\bar{R}_{{\rm e},x}\approx (N-1)b$ in the high-force limit, where lateral deflections of the chain are not significant. In addition, for a highly stretched chain, $b\approx b_{\rm max}$, and so $l_{\rm K}\approx b_{\rm max}$. Defining $\xi\equiv 1- \bar{b}/b_{\rm max}$ and noting $k_{\rm B}T=1$, it follows that $\xi \approx \sqrt{1/2b_{\rm max}}f^{-0.5}=0.75f^{-0.5}$, in excellent agreement with the results. This consistency of the stretching behavior at high force with the predictions of Marko and Siggia  was also observed in Ref.~\onlinecite{chen2024nonlinear} for polycatenane composed of flexible rings. 

\subsection{Wall-tethered polycatenane: Scaling with $N$}
\label{subsec:scaling}

\begin{figure}[!ht]
\begin{center}
\vspace*{0.1in}
\includegraphics[width=0.45\textwidth]{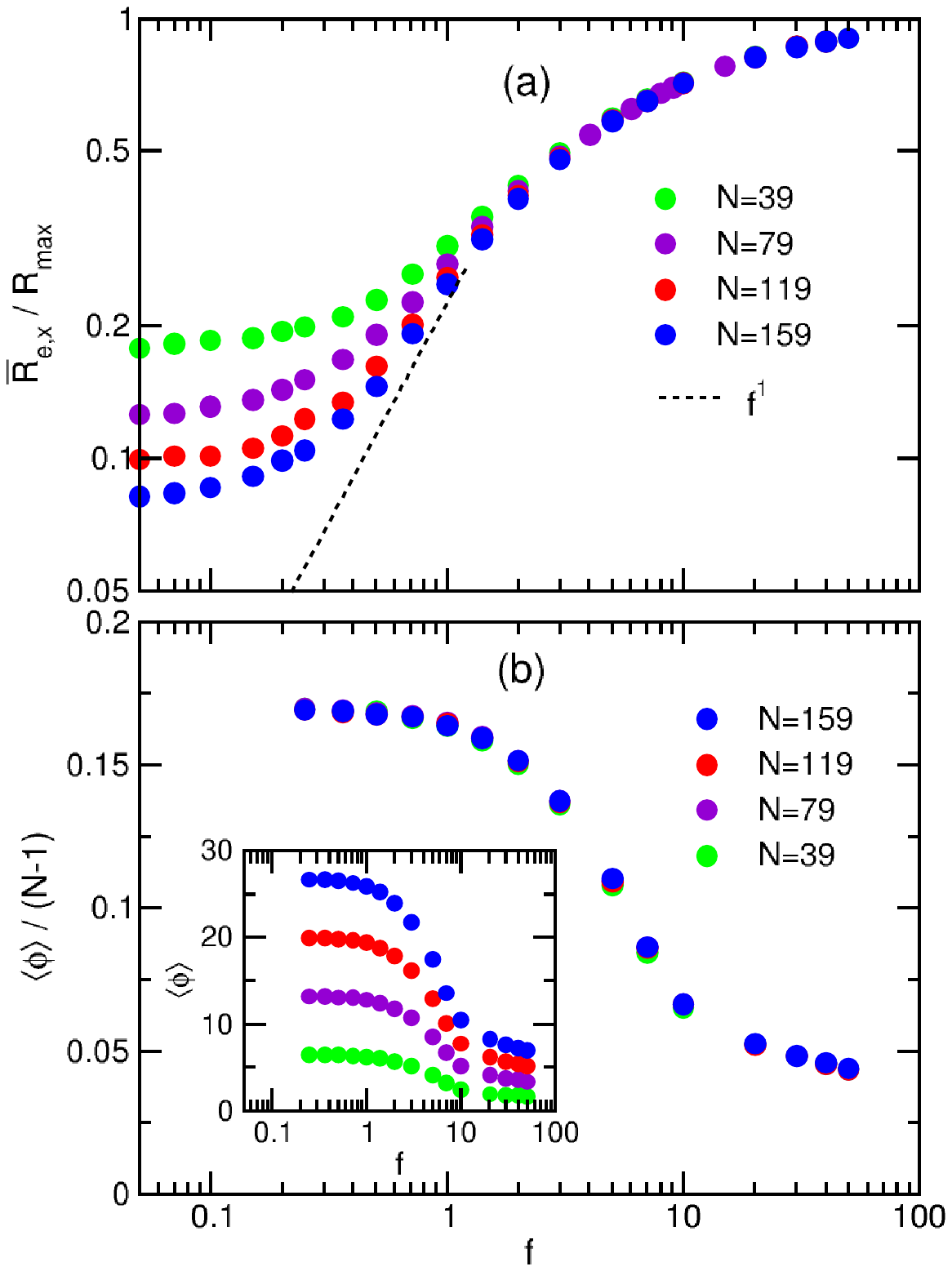}
\vspace*{0in}
\end{center}
\caption{(a) Scaled mean extension length vs force for a polycatenane chain confined between two hard walls. No torque is applied to the end rings. Results for various values of the chain length are shown. The dashed line shows the scaling expected for standard polymers in the weak-force limit in the absence of confinement. (b) Scaled mean twist angle, $\langle\phi\rangle/(N-1)$, vs extension force for polycatenane chains of various lengths. In each case, the applied torque is $\tau$=10. The inset shows the data for the unscaled twist angle. }
\label{fig:rex_phi_force_N}
\end{figure}

Having characterized the conformational behavior of stretched polycatenane in the absence of confinement, we next consider the case where the confining walls are present. As noted earlier, the walls are necessary for the chain to hold its supramolecular twist upon the application of a torque to the end rings. We first examine the case of no applied torque ($\tau=0$). Figure~\ref{fig:rex_phi_force_N}(a) shows the variation of the scaled extension length, $\bar{R}_{{\rm e},x}/R_{\rm max}$, vs force, $f$. Results are shown for four different chain lengths ($N$=39, 79, 119 and 159). As was the case for unconfined stretched polymers, the scaled extension length increases with increasing force and levels off asymptotically to the maximum extension length. In addition the scaled extension length for all chain lengths  converge for $f\gtrsim 1$. Unlike the case for unconfined chains, $\bar{R}_{{\rm e},x}$ does not converge at low force to power-law scaling $f^\alpha$ with an exponent of $\alpha\approx 1$. Instead, the extension length levels off asymptotically to a finite value as the force decreases. In addition, the scaled extension $\bar{R}_{{\rm e},x}/R_{\rm max}$ at any given force in this regime increases as the chain length decreases. This trend can be understood as follows. At low force and extension length, the walls are close together. As the chain is confined to the space between the walls, the conformational entropy of the chain will decrease significantly at these short inter-wall distances. In effect, this creates an entropic force pushing back on the walls, thus increasing the distance between the end rings. In the weak-force limit, it is expected that the end-ring distance (and inter-wall distance) will be $R_{{\rm e},x}\approx R_{\rm g}$, where $R_{\rm g}$ is the radius of gyration of the free chain. Since $R_{\rm g}\sim N^\nu$ and $R_{\rm max}\sim N$, it follows that $\bar{R}_{{\rm e},x}/R_{\rm max} \sim N^{\nu-1}\approx N^{-0.4}$, using $\nu\approx 0.6$. A fit of the scaled extension length at $f=0.05$ vs $N$ yields  $\bar{R}_{{\rm e},x}/R_{\rm max} \approx N^{-0.53\pm 0.04}$, in rough agreement with this simplistic prediction.  At higher forces, the polymer is more stretched out, leading to fewer interactions of rings with the walls and thus a negligible confinement effect. Consequently, the extension lengths for the confined systems converge to those for wall-less systems, for which the results for different chain lengths converge. 

Next we consider the scaling of the twist angle with respect to chain length for a polycatenane chain with a torque applied between the end rings. Figure~\ref{fig:rex_phi_force_N}(b) shows the variation of the mean twist angle per link, $\langle\phi\rangle/(N-1)$, vs the extension force, $f$, for chains of four different lengths, each subject to a torque of $\tau=10$. As expected, the torque causes the end rings to twist, leading to a non-zero $\langle\phi\rangle$. This twist angle decreases monotonically for increasing force, demonstrating that the stretching force and applied torque are opposed in their effect on chain twisting. Most notably, the results for $N$=39, 79, 119 and 159 overlap very closely over the entire range of force examined, with minor differences only barely visible in the case of the shortest chain. Thus, the twist angle is simply proportional to the total number of links, regardless of the extension force. This contrasts with the results for the scaled extension length in Fig.~\ref{fig:rex_phi_force_N}(a), where such scaling is present only at medium and higher forces.

\subsection{Wall-tethered polycatenane: Effects of varying force and torque}
\label{subsec:polycat_stretch_twist_confine}

\begin{figure*}[!ht]
\begin{center}
\includegraphics[width=0.9\linewidth]{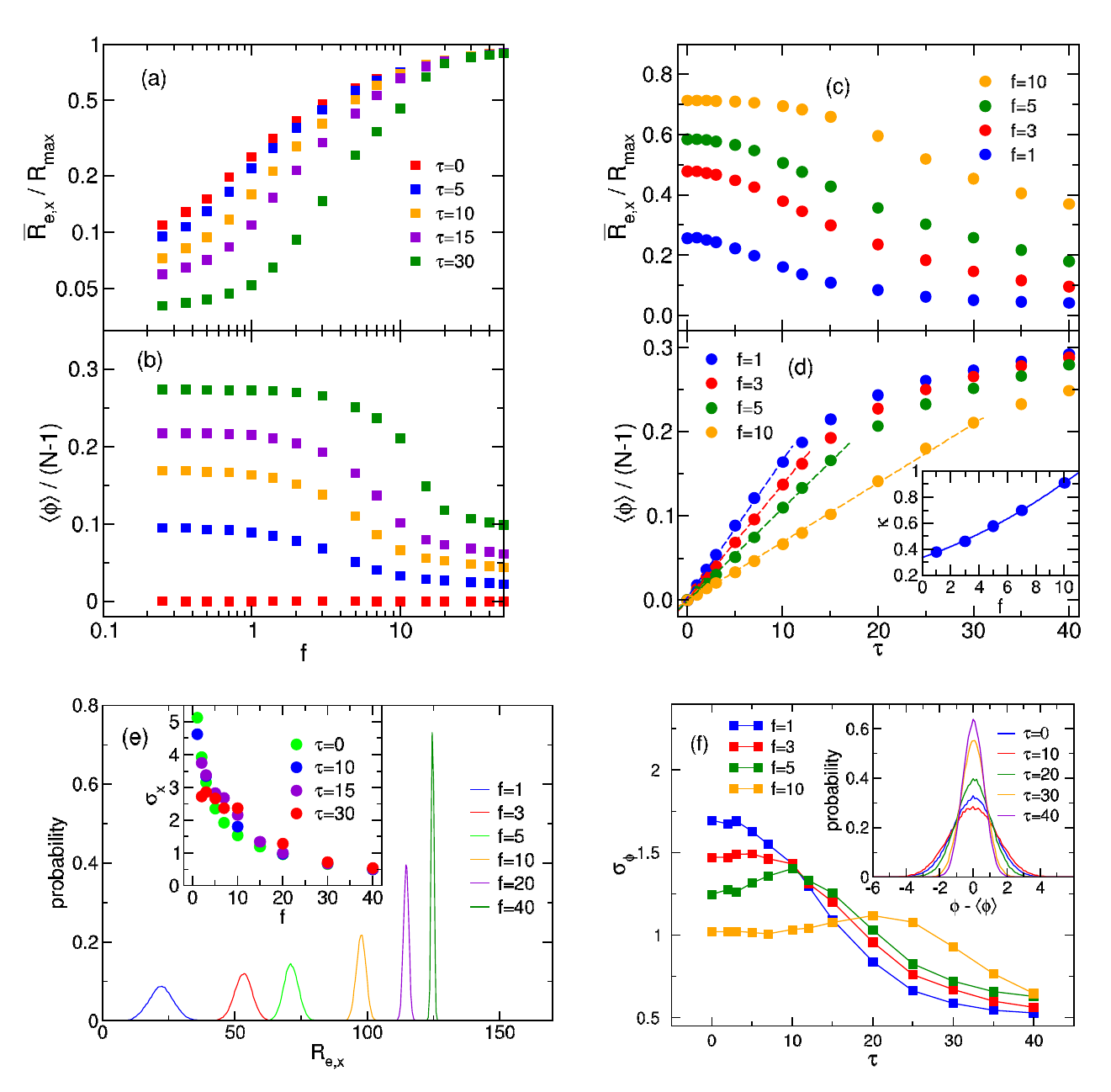}
\vspace*{-0.3in}
\end{center}
\caption{(a) Scaled mean extension length vs applied force for a polycatenane chain of length $N=159$.  Results are shown for various values of the applied torque.  (b) Mean end-ring twist angle per link vs extension force. (c) Scaled mean extension length vs applied torque.  Results are shown for various values of the extension force, $f$. (d) Mean end-ring twist angle per link vs applied torque. The dashed lines show linear fits to the data at low values of $\tau$. The inset shows the variation of torsional constant $\kappa$ with respect to applied force $f$, where $\kappa$ is defined by $\tau = \kappa\langle\phi\rangle$, and where $\kappa$ is calculated from the linear fits of $\langle\phi\rangle/(N-1)$ vs $\tau$ in the linear regime. (e) Probability distribution of extension length for an applied torque of $\tau=10$ and for various values of the extension force.  The inset shows the variation of $\sigma_x\equiv\sqrt{\langle R_{{\rm e},x}^2\rangle-\langle R_{{\rm e},x}\rangle^2}$ with respect to the extension force for various values of the applied torque. (f) Variation of $\sigma_\phi\equiv\sqrt{\langle\phi^2\rangle-\langle\phi\rangle^2}$ vs applied torque. Results for various values of the extension force are shown. The inset shows probability distributions for the twist angle for the case of $f=5$ for various values of the applied torque.}
\label{fig:rex_phi_force_tau}
\end{figure*}

Next, we carry out a systematic analysis of the combined effects of the force and torque on the extension length and twist angle for polycatenane chains of a fixed length of $N=159$. Figure~\ref{fig:rex_phi_force_tau} shows a collection of results for $\bar{R}_{{\rm e},x}$ and $\langle\phi\rangle$ upon variation of $f$ and $\tau$. Figure~\ref{fig:rex_phi_force_tau}(a) shows the variation of $\bar{R}_{{\rm e},x}/R_{\rm max}$ with $f$. Results for several different values of the applied torque are shown. As was evident for the case of $\tau=0$  in the results in Fig.~\ref{fig:rex_phi_force_N}(a), the extension length increases with increasing force for all values of $\tau$. Over most of the range of force considered, the extension length decreases with increasing torque. Thus, twisting the chain has the effect of pulling the two ends of the chain together, thus opposing the effect of the stretching force. For sufficiently high force, the results eventually do converge as the chain extension approaches its limiting value. The reduction in extension length with increasing $\tau$ (i.e., with increasing twist angle) is a result comparable to that observed in Ref.~\onlinecite{tubiana2022circular}, where an increase in the twist of circular polycatenanes was found to decrease their average size.

Figure~\ref{fig:rex_phi_force_tau}(b) shows the variation of the mean twist angle per per link, $\left\langle \phi \right\rangle/(N-1)$, vs the extension force, $f$ for the same values of $\tau$ as in (a). As expected, the mean twist angle is zero in the absence of a torque and increases as $\tau$ increases. For any given torque, the mean twist angle decreases monotonically with increasing force, again demonstrating the opposing effects of the torque and force on the chain configurations. For all values of torque considered, the decrease in $\langle\phi\rangle$ with force is pronounced in the range $f=1-10$ but then slows down at higher values of $f$. 

Figures~\ref{fig:rex_phi_force_tau}(c) and (d) show the variation of the scaled extension length and twist angle, respectively, with respect to torque for fixed stretching force. Results are shown for various values of the force. Consistent with Fig.~\ref{fig:rex_phi_force_tau}(a), $\bar{R}_{{\rm e},x}$ increases with increasing force and generally decreases with increasing torque. Notably, the rate of decrease in extension length with torque is slower for higher values of the force. For the lowest value of the force shown ($f=1$), the extension length appears to level off at high $\tau$, and it is likely that the same trend would be present for higher forces at values of $\tau$ beyond those considered here. Figure~\ref{fig:rex_phi_force_tau}(d) shows the variation of the mean twist angle per ring with $\tau$ for various values of the stretching force. As expected, $\langle\phi\rangle$ increases monotonically with $\tau$, though the rate of increase slows at higher values of $f$. For each value of the force, there is a range of torque over which the mean twist angle varies linearly with the torque. This linear range widens with increasing force, increasing from $\tau\approx 0-10$ for $f=1$ to $\tau\approx 0-30$ for $f=10$. We define the entropic torsional spring constant, $\kappa$, such that $\tau=\kappa\langle\phi\rangle$ in this regime and fit the data to obtain values of $\kappa$ for each force. The inset of Fig.~\ref{fig:rex_phi_force_tau}(d) shows the variation of $\kappa$ with $f$. The increase of $\kappa$ with $f$ shows that increasing the stretching force requires a larger torque to achieve a given mean twist angle. 

Figure~\ref{fig:rex_phi_force_tau}(e) shows probability distributions the extension length of a polycatenane chain of length $N$=159 for an applied torque of $\tau=10$ and for various values of the extension force. The inset shows the variation of $\sigma_x\equiv\sqrt{\langle R_{{\rm e},x}^2\rangle-\langle R_{{\rm e},x}\rangle^2}$ with respect to the extension force. Results for various values of the applied torque are shown. Generally, the variance in $R_{{\rm e},x}$ decreases with increasing force with values that are roughly independent of the applied torque. 
Figure~\ref{fig:rex_phi_force_tau}(f) shows the variation of $\sigma_\phi\equiv\sqrt{\langle\phi^2\rangle-\langle\phi\rangle^2}$ with the respect to applied torque for various values of the extension force. The inset shows probability distributions for the twist angle for the case of $f=5$. Results for various values of the applied torque are shown. The trends for $\sigma_\phi$ are more complex than those for $\sigma_x$, with the variation of $\sigma_\phi$ with $\tau$ strongly affected by the value of the stretching force. At low values of $f$, the variance decreases monotonically with $\tau$, but at higher $f$, there is a local maximum whose position shifts to higher $\tau$ with increasing force. 

\subsection{Ribbons}
\label{subsec:ribbons}

\begin{figure*}[!ht]
\begin{center}
\vspace*{0.0in}
\includegraphics[width=0.9\textwidth]{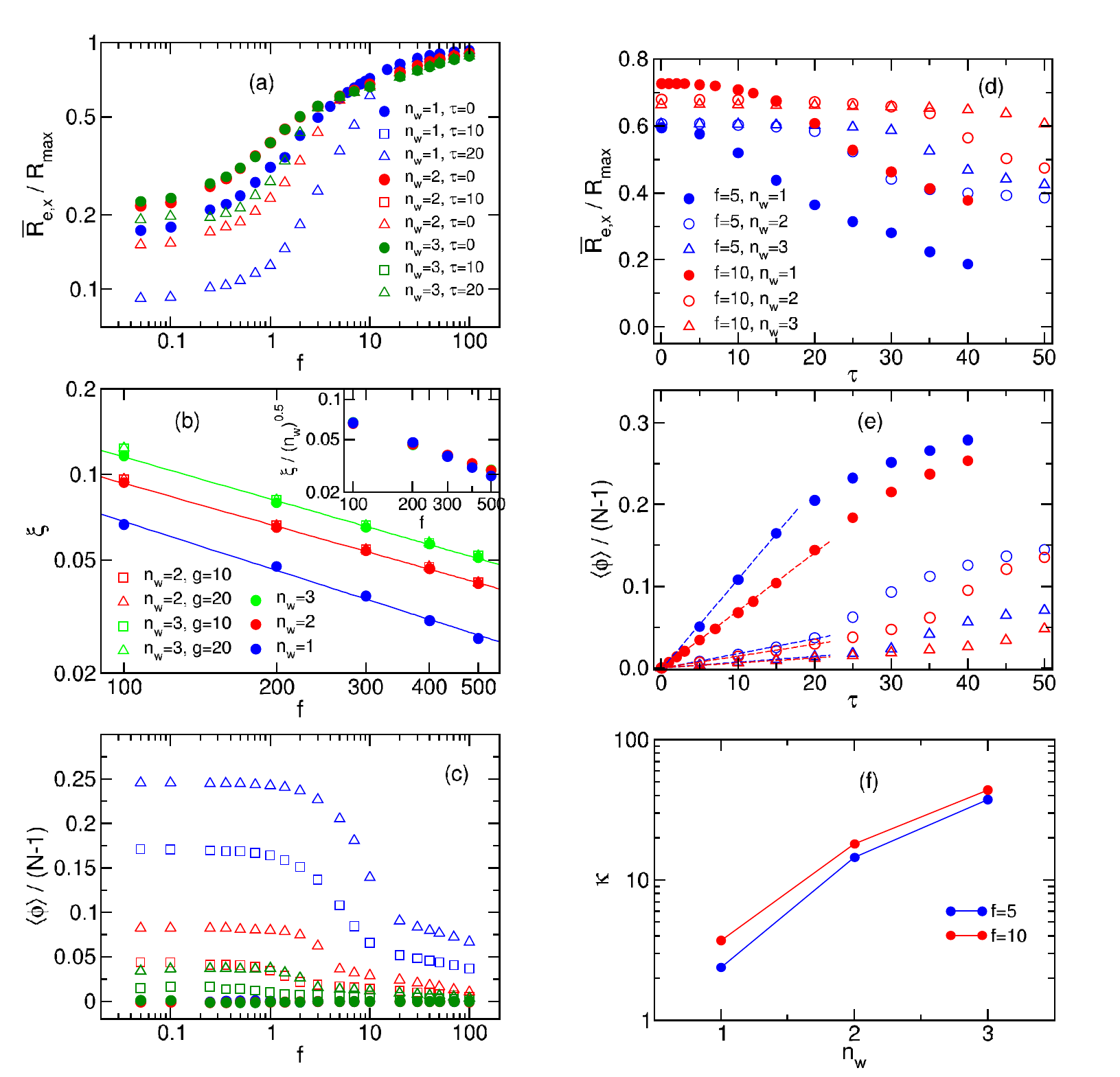}
\vspace*{-0.2in}
\end{center}
\caption{(a) Scaled mean extension length vs applied force for ribbons of length  $N=39$.  Results are shown for various values of the applied torque, $\tau$. 
(b) Variation of $\xi\equiv 1 - \bar{R}_{{\rm e},x}/R_{\rm max}$ vs force in the high-force limit for ribbons of length $N=41$ and various widths for $\tau=0$. The solid lines show fits to the power function $\xi=cf^{-\alpha}$. The fits yield values of $\alpha=0.57\pm 0.03$ for $n_{\rm w}=1$, $\alpha=0.50\pm 0.01$ for $n_{\rm w}=2$, and  $\alpha=0.51\pm 0.01$ for $n_{\rm w}=3$. Overlaid on these data are those for ribbons with gaps with $g=10$ and $g=20$ for $n_{\rm w}=2$ and $n_{\rm w}=3$. The inset shows $\xi/\sqrt{n_{\rm w}}$ vs force for gap-less ribbons of width $n_{\rm w}$=1, 2 and 3.
(c) As in (a), except mean end-ring twist angle per link vs applied applied force.
 The legend is the same as in (a).
 (Note: to reduce visual clutter, the results for $\tau$=10 are displayed in (c), but not in (a).) 
(d) Mean extension length per ring vs applied torque for ribbons of length  $N=39$.  Results are shown for various values of the extension force, $f$. 
(e) As in (d), except mean end-ring twist angle per ring vs applied torque. The dashed lines show linear fits to the data with low values of $\tau$.
(f) Variation of torsional spring constant $\kappa$ vs ribbon width, $n_{\rm w}$. The torsional constant is defined by the relation $\langle\phi\rangle = \kappa\tau$, where $\kappa$ is calculated from the fits of $\langle\phi\rangle$ vs $\tau$ in the linear regime. 
 }
\label{fig:rex_phi_force_N2}
\end{figure*}

Next we examine the effects of stretching and twisting catenane ribbons, illustrated in Figs.~\ref{fig:illustration1}(b) and (c). The effects of varying the applied force and torque on the extension length and the mean twist angle are illustrated in Fig.~\ref{fig:rex_phi_force_N2}. Results are shown for ribbons of length $N=39$ and widths of $n_{\rm w}=$1, 2 and 3, each for torques of $\tau$=0, 10 and 20. Figure~\ref{fig:rex_phi_force_N2}(a) shows the variation of the scaled extension, $\bar{R}_{{\rm e},x}/R_{\rm max}$, with force. 
(Note that results for $\tau=10$ are omitted in Fig~\ref{fig:rex_phi_force_N2}(a) to reduce the clutter of that part of the figure.) As observed previously for polycatenane ($n_{\rm w}=1$),  the extension length increases montonically with increasing $f$ for ribbons with $n_{\rm w}=2$ and $n_{\rm w}=3$, as well, in each case gradually levelling off at high force as the ribbons approach the maximum extension imposed by the linking constraints. Likewise, the behavior of the mean twist angle as the force is varied for ribbons with $n_{\rm w}=2$ and $n_{\rm w}=3$ is qualitatively comparable to that for polycatenane. 

The effects of varying the ribbon width, $n_{\rm w}$,  are pronounced. At $\tau=0$, the extension length of the ribbons with $n_{\rm w}=2$ and $n_{\rm w}=3$ are very close to each other and noticeably different from that for polycatenane ($n_{\rm w}=1$). At a low force of $f \lesssim 5$, the ribbon extension length length is significantly higher than that for polycatenane, while at higher force $f\gtrsim 5$ the trend reverses, and the ribbon extension is lower than that of polycatenane. The origin of this crossover is straightforward. As noted in Fig.~\ref{fig:rg_vs_n1}, the average size of a free chain is appreciably larger for $n_{\rm w}=2$ and $n_{\rm w}=3$ than for $n_{\rm w}=1$. Consequently, when the ribbons are stretched, a smaller force would be required than for polycatenane to achieve a given extension, as it will need to be stretched over smaller displacement. This effect is likely compounded by the confinement imposed by the walls: the entropic pressure pushing the walls (and thus the end rings) apart is expected to be greater for the chains with the larger natural size (i.e. $n_{\rm w}\geq 2$ chain), thus further increasing the end-ring separation for a given value of $f$. At larger force, the chains are highly stretched and rarely interact with the walls, and so this confinement effect is negligible. More significant is the apparent increase in an entropic resistance to increasing the extension as the ribbon width increases. In short, catenane ribbons are easier than polycatenane to stretch at short distance but harder to do so when they are already highly extended.

Figure~\ref{fig:rex_phi_force_N2}(b) shows the effects of varying the force on the extension length in the high-force limit of $f\geq 100$. Results are shown for $\tau=0$ for chains of width $n_{\rm w}$=1, 2 and 3. The graph shows that the variation of $\xi\equiv 1 - \bar{R}_{{\rm e},x}/R_{\rm max}$ with respect to $f$  exhibits power-law behavior in this regime as was the case for unconfined polycatenane, illustrated in the inset of Fig.~\ref{fig:rex_Q_freeends}(b). Fitting the data to the function $\xi=cf^{-\alpha}$ yields scaling exponents of  $\alpha=0.57\pm 0.03$ for $n_{\rm w}=1$, $\alpha=0.50\pm 0.01$ for $n_{\rm w}=2$, and  $\alpha=0.51\pm 0.01$ for $n_{\rm w}=3$. As noted earlier in Sec.~\ref{subsec:polycat_stretch}, the Marko-Siggia model for wormlike chains\cite{marko1995stretching} predicts a high-force-limit scaling of $\xi  \approx \sqrt{k_{\rm B}T/2fl_{\rm K}}\propto f^{-0.5}$. Thus, the results for the scaling with respect to force for catenane chains are consistent with the Marko-Siggia model in this limit, with a slightly greater deviance for the case of polycatenane ($n_{\rm w}=1$) than for the two ribbons. However, the observed increase in $\xi$ with ribbon width is not accounted for by this single-chain model. As $n_{\rm w}$ increases, the Kuhn length is expected to increase, which should lead to a decrease in $\xi$ with $n_{\rm w}$. To investigate this effect further, we introduce gaps in the side links of the chains in the $n_{\rm w}=2$ and $n_{\rm w}=3$ ribbons. The results for gaps of $g=10$ and $g=20$ are overlaid on the other data and clearly show negligible difference between ribbons with gaps and the gap-less ribbons. Note that $g=20$ corresponds to completely disconnected polycatenane chains. Thus, at high force the stretching behavior of ribbons of width $n_{\rm w}$ is the same as that for $n_{\rm w}$ disconnected polycatenane chains, with the side linking between the chains having negligible effect. Increasing the number of stretched polycatenane chains increases the entropic resistance to the stretching force. Thus, a greater force is required to achieve a given extension, which accounts for the observed increase in $\xi$ with $n_{\rm w}$. To quantify this effect, first note that the high-force limit of the Marko-Siggia result for a single chain can be written: $f_1 = (k_{\rm B}T/2l_{\rm K})\xi^{-2}$. For $n_{\rm w}$ chains at the same extension (and thus the same $\xi$), $f_{n_{\rm w}} = n_{\rm w}f_1$. It follows that when single-chain and $n_{\rm w}$-chain systems are stretched with the same force, the extensions are related by $\xi_{n_{\rm w}}/\xi_1 = \sqrt{n_{\rm w}}$. Thus, $\xi_{n_{\rm w}}/\sqrt{n_{\rm w}}$ is predicted to be a universal curve for all ribbon widths. The variation of  inset of $\xi_{n_{\rm w}}/\sqrt{n_{\rm w}}$ with force is shown the inset of Fig.~\ref{fig:rex_phi_force_N2}(b) for the gap-less ribbons of width $n_{\rm w}$=1, 2 and 3. With some minor deviations at high force for $n_{\rm w}=1$, the data collapse onto a single curve, in accord with the predictions of the theoretical model.

Consider now the effect of applying a torque on the extension length. As is evident in Fig.~\ref{fig:rex_phi_force_N2}(a), the force-dependence of the chain extension is significantly impacted. In the case of $\tau=20$, the extension length decreases significantly for all chain widths for lower values of $f$, but is minimally impacted when subject to a higher stretching force. The decrease in $\bar{R}_{{\rm e},x}$ with increasing torque is most pronounced for polycatenane chains ($n_{\rm w}=1$) and becomes less pronounced as the chain width $n_{\rm w}$ increases. In addition, the force range within which imposing a torque has this effect on the chain extension is reduced as the chain width increases. As a quantitative measure of this effect, the curve for $\tau=20$ converges on the curve for $\tau=0$ at $f \approx 15$ for $n_{\rm w}=1$, at $f\approx 5$ for $n_{\rm w}=2$, and $f\approx 2$ for $n_{\rm w}=3$. 

Figure~\ref{fig:rex_phi_force_N2}(c) shows the variation of the scaled mean twist angle with respect to force for various values of $\tau$ and chain width. As expected, the mean twist angle is zero for all chains when $\tau=0$. For $\tau>0$, the variation of $\langle\phi\rangle$ with $f$ follows the same qualitative behavior discussed previously for polycatenane chains in Fig.~\ref{fig:rex_phi_force_tau} in the cases of $n_{\rm w}=2$ and $n_{\rm w}=3$. The variation of the scaled mean twist angle with respect to force shown in  Fig.~\ref{fig:rex_phi_force_N2}(c) likewise shows a pronounced effect when the chain width increases. At a given torque, the mean twist angle decreases significantly when $n_{\rm w}$ increases. Thus, increasing the ribbon width effectively increases the entropic torsional stiffness of the chains.

Figure~\ref{fig:rex_phi_force_N2}(d) shows the variation of the chain extension with $\tau$ for various values of the stretching force and the chain width. In the case of $f=5$, the response is noticeably different for polycatenane than for $n_{\rm w}\geq 2$. In the former case, increasing the torque shortens the extension length monotonically for all $\tau\geq 0$. By contrast in the case of ribbons the extension length is invariant with respect to increasing torque until it reaches a threshold value ($\tau\approx 20$ for $n_{\rm w}=2$ and $\tau\approx 30$ for $n_{\rm w}=3$), beyond which $\bar{R}_{{\rm e},x}$ begins to decrease with increasing $\tau$. Qualitatively similar behavior is seen for a higher extension force of $f=10$. The main quantitative effect of increasing the stretching force is to broaden the range of $\tau$ over which the ribbon extension length remains invariant to the torque. Also of note for $f=10$ is the crossing of the curve for $n_{\rm w}=1$ with those for $n_{\rm w}=2$ and $n_{\rm w}=3$. At this higher value of the force, polycatenane is more stretched at lower torque than the ribbons, but is less stretched for higher values of $\tau$. 

Figure~\ref{fig:rex_phi_force_N2}(e) shows the variation of the scaled mean twist angle with respect to torque. The increase in $\langle\phi\rangle$ with $\tau$ for all systems, previously noted above for Fig.~\ref{fig:rex_phi_force_N2}(c), is very clear, as is the resistance to end-ring twisting as the ribbon width increases. As in Fig.~\ref{fig:rex_phi_force_tau}(d) for the case of polycatenane, there is a low-torque regime within which there is a linear variation of the twist angle with $\tau$. In addition, as in that case, the range over which the variation is linear widens with increasing force.
Linear fits of the data in this regime can be used to extract the entropic torsional spring constant $\kappa$ (defined by the relation $\langle\phi\rangle = \kappa\tau$). Figure~\ref{fig:rex_phi_force_N2}(f)   shows the variation of $\kappa$ with $n_{\rm w}$ for the two values of the force. For each $f$, $\kappa$ increases by over an order of magnitude when going from polycatenane ($\kappa\approx 0.6$ for $f=5$ and $\kappa\approx 0.9$ for $f=10$) to ribbons with $n_{\rm w}\geq 2$. Thus, as noted for the results in Fig.~\ref{fig:rex_phi_force_N2}(c), increasing the ribbon width significantly increases its torsional stiffness.

\subsection{Ribbons with gaps}
\label{subsec:ribbons_gaps}

\begin{figure*}[!ht]
\begin{center}
\vspace*{0.0in}
\includegraphics[width=0.9\textwidth]{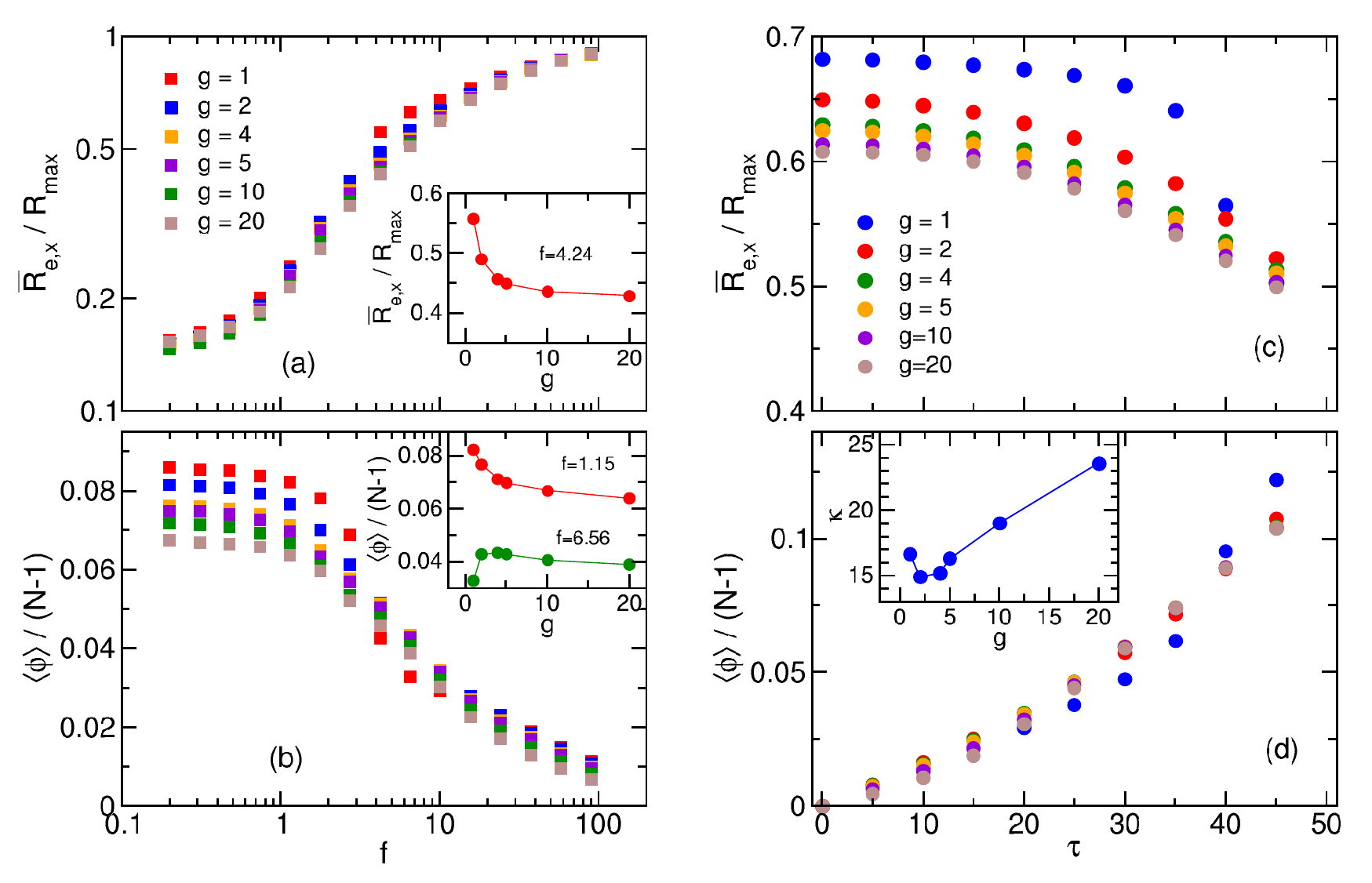}
\vspace*{-0.2in}
\end{center}
\caption{Effects of the varying the gap length $g$ on the stretching and twisting of ribbons of length $N=41$ and width $n_{\rm w}=2$.  (a) Scaled mean extension length vs applied force for ribbons with an applied torque of $\tau=20$.  Results are shown for various values of $g$. The inset shows extension length vs $g$ for $f=4.24$. (b) As in (a), except mean end-ring twist angle per link vs $g$. The inset shows the variation of $\langle\phi\rangle/(N-1)$ with $g$ for $f=1.15$ and $f=6.56$.  (c) Scaled mean extension length vs applied torque for ribbons with an applied force of $f=10$.  Results are shown for various values of $g$. (d) As in (c), except mean end-ring twist angle per link vs applied torque. The inset shows the variation of torsional spring constant $\kappa$ vs gap length, $g$. The torsional constant is defined through the relation $\langle\phi\rangle = \kappa\tau$, where $\kappa$ is calculated from the fits of $\langle\phi\rangle$ vs $\tau$ in the linear regime. 
}
\label{fig:rex_phi_gaps_n2_all}
\end{figure*}

As a final aspect of catenane elasticity to explore in this study, we examine the effects of introducing gaps in the cross links of the ribbons, as illustrated in Fig.~\ref{fig:illustration1}(d) and (e). Results for ribbons of length $N=41$ and width $n_{\rm w}=2$ are show in Fig.~\ref{fig:rex_phi_gaps_n2_all}. Figures~\ref{fig:rex_phi_gaps_n2_all}(a) and (b) show the variation of the scaled extension and twist angle per link, respectively, with respect to force at a fixed torque of $\tau=20$. At high and low $f$, the extension length for various $g$ roughly converge, though there is a detectable decrease in $\bar{R}_{{\rm e},x}$ with increasing $g$ . Only at intermediate values of force near $f\approx 3-10$ is the extension length for a gap-less ribbon ($g=1$) noticeably larger than that for the $g>1$ ribbons. This is evident in the inset, which shows ribbon extension vs $g$ for $f=4.24$. There is a 23\% reduction in the extension length between the gap-less ribbon ($g=1$) and the case where the ribbon has been divided into two disconnected chains ($g=20$). Figure~\ref{fig:rex_phi_gaps_n2_all}(b) reveals similar anomalous behavior for the mean twist angle.  At high and low force, the twist angle tends to decrease slightly with increasing $g$. However, at intermediate force, the twist angle for the gap-less ribbon ($g=1$) is somewhat lower that the $g>1$ ribbons. The inset highlights this contrast in plotting the twist angle vs $g$ for a low force ($f=1.15$) and an intermediate force ($f=6.56$). Overall, the effects on chain extension of introducing gaps are small, indicating that the response to varying the force is dominated by the simple fact that there are two polycatenane chains, with the number of side links playing a minor role.

Figures~\ref{fig:rex_phi_gaps_n2_all}(c) and (d) show the variation of the scaled extension and twist angle per link, respectively, with respect to torque at a fixed force of $f=10$. Increasing the gap length tends to reduce the chain extension over most of the range of $\tau$ considered except at a high value of $\tau=50$, by which point the results appear to converge. As an example, at $\tau=20$, the extension length reduces by 12\% in going from gap-less ($g=1$) to side-link-less ($g=20$) ribbons. Figure~\ref{fig:rex_phi_gaps_n2_all}(d) appears to show a relatively small effect of varying $g$ on the dependence of the twist angle on the torque, with most of the data overlapping except for those for $g=1$ at $\tau \gtrsim 25$. However, somewhat more interesting behavior is revealed by carrying out fits of the data in a low-$\tau$ linear regime to extract a torsional spring constant, $\kappa$, as was done for the data in Figs.~\ref{fig:rex_phi_force_tau}(d) and \ref{fig:rex_phi_force_N2}(d). (To avoid unnecessary clutter in the figure, the lines of best fit are omitted.) The inset shows the variation of $\kappa$ with respect to $g$. Initially, the torsional stiffness decreases slightly with increasing $g$ in accord with the expectation that decreasing the connectivity of a catenane will make it easier to deform (as in the case of kinetoplasts, where the bending rigidity is reduced when the connectivity of the DNA rings is reduced\cite{ramakrishnan2024single}). However, at higher $g$ the torsional constant significantly {\it increases} with $g$. Specifically, there is a 42\% increase in $\kappa$ between the $g=1$ (gap-less) and $g=20$ (side-link-less) extremes. This effect can be understood as follows. For the $N=41$ chains employed in the simulations, $g=10$ and $g=20$ correspond to to a single side-ring link and zero side-ring links, respectively. In the latter case, the rings are completely disconnected and interact solely through excluded-volume interactions. In this limit, the conformational entropy is much higher than for when the chains are tightly bound to one another. Twisting the chains by applying a torque will cause the chains to wind around each other, which in turn will suppress the independence of the lateral fluctuations of the individual chains. This will lead to a significant reduction in the configurational entropy that shows up as higher values of the (entropic) torsional stiffness relative to the case where the chains are more tightly bound. 

\begin{figure}[!ht]
\begin{center}
\vspace*{0.0in}
\includegraphics[width=0.42\textwidth]{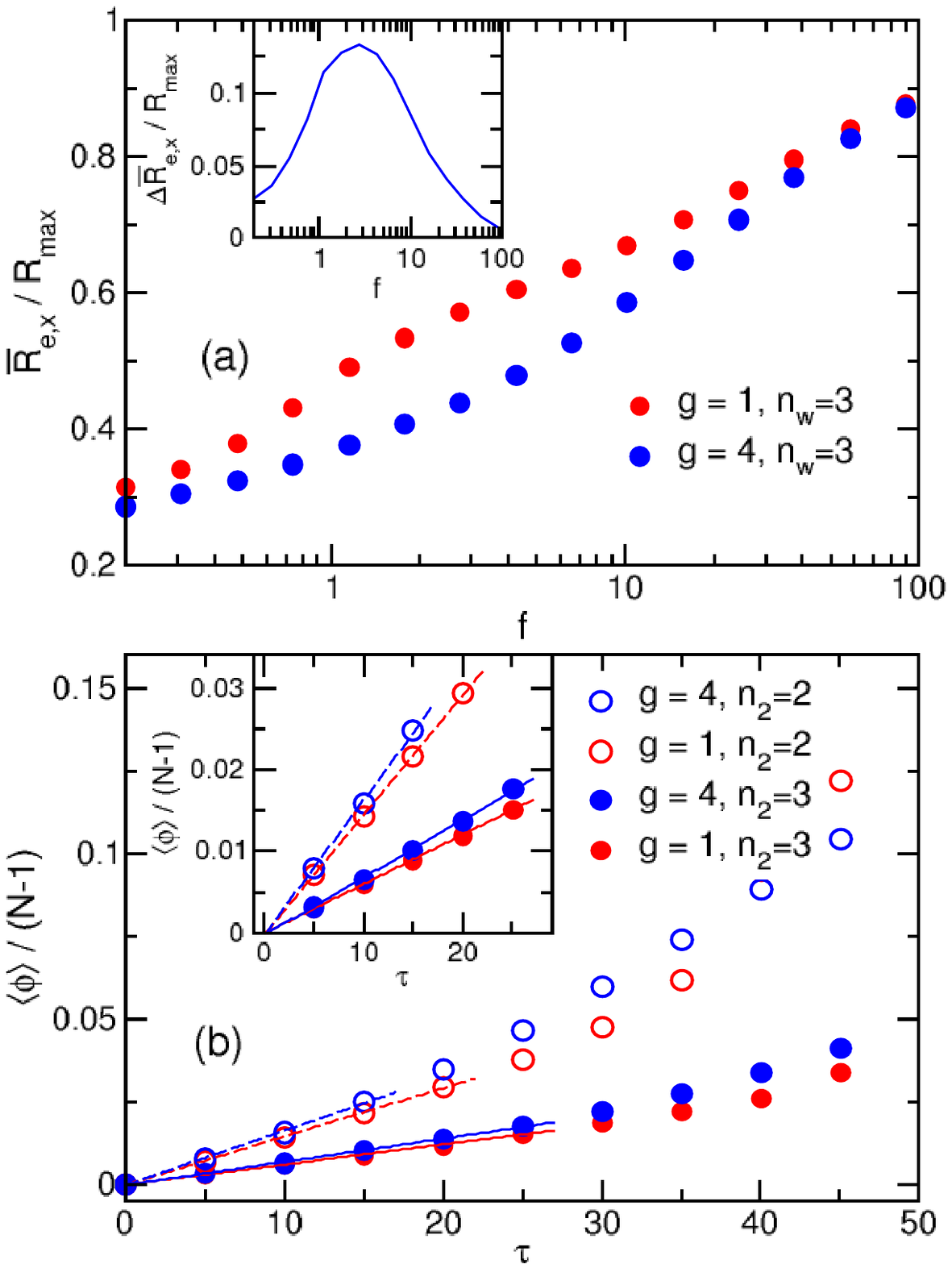}
\vspace*{-0.2in}
\end{center}
\caption{Effects of the varying the gap length $g$ on the stretching and twisting of ribbons of length $N=41$ and width $n_{\rm w}=3$.  (a) Scaled mean extension length vs applied force for ribbons with no applied torque.  Results are shown for $g=1$ and $g=4$. The inset shows the difference between the two sets of results. (b) Scaled twist angle vs torque for ribbons stretched with a force of $f=10$. For comparison, results for $n_{\rm w}=2$ ribbons are overlaid on the results for $n_{\rm w}=3$. The solid and dashed lines are fits to the data in the low-$\tau$ linear regime. The inset shows a close-up of the fits and data in the linear regime. The fits yield the following values of the torsional constant: For $n_{\rm w}=3$, $\kappa=34.9$ for $g=1$ and $\kappa=41.3$ for $g=4$. For $n_{\rm w}=2$, $\kappa=14.8$ for $g=1$ and $\kappa=16.8$ for $g=4$.}
\label{fig:rex_phi_force_torque_n3}
\end{figure}

Figure~\ref{fig:rex_phi_force_torque_n3} shows the effects of gaps on wider ribbons of width $n_{\rm w}=3$. Figure~\ref{fig:rex_phi_force_torque_n3}(a) compares the force dependence of the extension length at zero torque on ribbons with no gaps ($g=1$) and with gaps of length $g=4$. For all values of $f$ examined, the gaps have the effect of reducing the extension length. As is especially clear from the inset of the figure, the difference is greatest in the intermediate force regime, consistent with the trends in Fig.~\ref{fig:rex_phi_gaps_n2_all}(a) for $n_{\rm w}=2$. Figure~\ref{fig:rex_phi_force_torque_n3}(b) shows the variation of the twist angle per link with respect to torque for a fixed force of $f=10$. Data for ribbons of $n_{\rm w}=2$ are also included in the graph for ease of comparison. In each case, results for gap-less ($g=1$) ribbons and those with $g=4$ are shown. As is the case for the narrow ribbons, the introducing gaps in the side links has the effect of  slightly increasing the twist angle at every value of $f$. Using the fits of the data in the low-$\tau$ linear region to extract the torsional spring constant $\kappa$, we find that the gaps decrease $\kappa$ from $\kappa=41.3$ for $g=1$ to $\kappa=34.9$ to $g=4$, an 18\% decrease. For comparison, in the case of $n_{\rm w}=2$, there is a 14\%  decrease from $\kappa=16.8$ for $g=1$ to $\kappa=14.8$ for $g=4$. Thus, the decrease  in torsional stiffness with increasing $g$  for low values of $g$ and for $n_{\rm w}=3$ is consistent with the trend observed for $n_{\rm w}=2$.

\vspace*{0.25in}
\section{Conclusions}
\label{sec:conclusions}

In this study, we have used MC simulations to characterize the elastic properties of catenane chains. While previous studies have focussed on polycatenane chain extension induced by an external force applied to the end rings, in this work we have also examined the torsional elastic properties by applying a torque to the end rings. These simulations are intended to mimic single-molecule force spectroscopy experiments employing AOTs, which have been previously used to study the elastic properties of biomolecules such as DNA. Our study also examined the behavior of catenane ribbons and the effects of varying the degree of connectivity of the polycatenane chains that are linked together to form the ribbon. Since a large number of time-consuming simulations were required to carry out this work, we chose to use rigid-ring catenanes to maximize computational efficiency, leaving the more physically realistic case of flexible-ring and semi-flexible-ring chains to a future study. As most other simulation studies of catenanes have employed (semi)flexible-ring models, we began our work by first characterizing the scaling properties of rigid-ring free polycatenanes and ribbons, as well as the extensional behavior of polycatenane in the absence of torque. Comparison of our results to those of previous studies provided a means to elucidate the effects of ring rigidity on the chain elasticity. 

We find that the scaling of the average size of free rigid-ring polycatenane and catenane ribbons with chain length $N$  exhibits power-law scaling with an exponent somewhat larger than the Flory exponent expected for self-avoiding chains. In part, this difference is attributed to finite-size effects, but in the case of the widest ribbon ($n_{\rm w}=3$) the noticeably higher exponent may arise from the predicted behavior of ribbon-like chains with a high stiffness in the ``turning'' and ``twisting'' modes.\cite{michaels2023conformational} The increase in the average size of the chains with ribbon width appears to be mainly due to an enhancement in excluded-volume interactions rather than the observed increase in chain stiffness. The behavior of polycatenane to an external force absent torque or confinement was consistent with previous studies using flexible-ring polycatenane.\cite{chen2023topological, chen2024nonlinear} As in those studies, we observe that the elastic modulus scales with force as $E\propto f^1$ (equivalently, $R_{{\rm e},x}$ varies linearly with respect to $ \ln(f)$) in an intermediate-force regime. In addition, we observe linear scaling at low force and scaling consistent with the Marko-Siggia prediction at high force. However, we do not observe the ``stress-softening'' sub-regime at intermediate forces observed by Chen {\it et al.} in Ref.~\onlinecite{chen2024nonlinear}, possibly a result of using rigid rings thereby preventing an oblate-to-prolate ring-shape transition that was associated with that regime. 

Application of a torque to the end rings of a stretched polycatenane chain had the effect of twisting the chain while simultaneously shortening its extension length. At low torque, the chain twist angle is proportional to the torque, and the associated torsional spring constant $\kappa$, itself a measure of torsional rigidity, was found to increase as the elongational force was increased. Similar behavior was also observed for catenane ribbons. In this latter case, the torsional stiffness increased dramatically with increasing ribbon width. In addition, we find that ribbons tend to be easier to stretch than polycatenane at low force and harder to stretch at high force. Interestingly, ribbons are more resistant to torque-induced shortening than polycatenane, a tendency that is enhanced by widening the ribbon. Reducing the lateral connectivity of the polycatenane chains linked to form the ribbons slightly reduced the extension  of the chains, thus increasing the extensional stiffness of the chains. In addition, the presence of such link gaps also tended to slightly decrease the torsional stiffness of the ribbons for small gaps, but significantly {\it increase} the torsional stiffness for larger gaps. The latter behavior contrasts with the observed significant reduction in the bending rigidity of 2D linked-ring kinetoplasts (admittedly, a rather different physical system) upon removal of some DNA rings using restriction enzymes.\cite{diggines2024multiscale, ramakrishnan2024single} 

In future work, we will extend the present study and examine the effects of ring stiffness on the stretching and torsional elasticity of  catenane chains by employing semi-flexible rings, which better describe the catenane systems studied experimentally  than do rigid-ring chains. Preliminary calculations suggest that variation of ring bending rigidity has a pronounced effect on catenane elastic behavior. We hope that experiments using single-molecule force spectroscopy methods will eventually be applied to study catenane chains in order to test the predictions of our simulations. 

\begin{acknowledgments}
This work was supported by the Natural Sciences and Engineering Research Council of Canada (NSERC). We are grateful to the Digital Research Alliance of Canada and the Atlantic Computational Excellence Network (ACEnet) for use of their computational resources. We would like to thank Sheldon Opps for helpful discussions.
\end{acknowledgments}


%

\end{document}